\documentclass[arps,prd,onecolumn,showpacs,nofootinbib]{revtex4}

\usepackage{epsfig}
\usepackage{epstopdf}
\input{epsf.sty}
\usepackage{epsf}
\usepackage{color}
\newcommand{\ba}{\begin{eqnarray}}
\newcommand{\ea}{\end{eqnarray}}
\newcommand{\be}{\begin{equation}}
\newcommand{\ee}{\end{equation}}

\newcommand{\vk}{{\bf k}}
\newcommand{\vx}{{\bf x}}

\newcommand{\hOMpc}{\,\textrm{h}/\textrm{Mpc}}
\newcommand{\MpcOh}{\,\textrm{Mpc}/\textrm{h}}

\def\thetaB{\mbox{\boldmath$\theta$}}

\def\justO{\mathcal{O}}

\newcommand\lsim{\mathrel{\rlap{\lower4pt\hbox{\hskip1pt$\sim$}}
        \raise1pt\hbox{$<$}}}
\newcommand\gsim{\mathrel{\rlap{\lower4pt\hbox{\hskip1pt$\sim$}}
        \raise1pt\hbox{$>$}}}

\input{epsf}        
\begin{document}
\title{Gravitational Lensing as Signal and Noise in Lyman-alpha Forest Measurements
}
\author{Marilena LoVerde$^{1,2}$, Stefanos Marnerides$^{2}$, Lam Hui$^{1,2,3}$,  Brice M\'{e}nard$^4$, Adam Lidz$^{5,6}$,}
\affiliation{$^{1}$Institute for Advanced Study, Princeton, NJ 08540 \\
$^{2}$ISCAP and
Department of Physics, Columbia University, New York, NY 10027\\
$^{3}$CCPP and Department of Physics, New York University, NY 10003\\
$^{4}$Canadian Institute for Theoretical Astrophysics, Toronto, Ontario M5S 3H8, Canada\\
$^{5}$Center for Astrophysics, Harvard University, Cambridge, MA 02138\\ 
$^{6}$Department of Physics and Astronomy, University of Pennsylvania, Philadelphia, PA 19104\\
marilena@ias.edu, stefanos@phys.columbia.edu, lhui@astro.columbia.edu, menard@cita.utoronto.ca, alidz@sas.upenn.edu}
\date{\today}
\begin{abstract}

In Lyman-$\alpha$ forest measurements it is generally assumed that quasars are mere background light sources which are uncorrelated with the forest. Gravitational lensing of the quasars violates this assumption. This effect leads to a measurement bias, but more interestingly it provides a valuable signal.  The lensing signal can be extracted by correlating quasar magnitudes with the flux power spectrum and with the flux decrement.  These correlations will be challenging to measure but their detection provides a \emph{direct} measure of how features in the Lyman-$\alpha$ forest trace the underlying mass density field. Observing them will test the fundamental hypothesis that fluctuations in the forest are predominantly driven by fluctuations in mass, rather than in the ionizing background, helium reionization or winds.  We discuss ways to disentangle the lensing signal from other sources of such correlations, including dust, continuum and background residuals. The lensing-induced measurement bias arises from sample selection: one preferentially collects spectra of magnified quasars which are behind overdense regions. This measurement bias is $\sim 0.1-1\%$ for the flux power spectrum, optical depth and the flux probability distribution. Since the effect is systematic, quantities such as the amplitude of the flux power spectrum averaged across scales should be interpreted with care. 

\end{abstract}


\maketitle
\section{Introduction}
\label{Intro}

Light from distant galaxies and quasars is gravitationally lensed by mass along the line of sight. For a flux-limited survey of quasars, lensing magnification biases the observed number density \cite{Turner:1984ch,Fugmann:1988,Narayan:1989}. For example, a positive mass fluctuation along the line of sight can increase the apparent sky area and the observed flux. The geometrical area increase decreases the quasar number density, while the flux increase promotes intrinsically faint objects above the magnitude threshold increasing the source density.  Together these effects introduce a correction to the quasar number density called magnification bias. Detections of the magnification of distant quasars by low-redshift galaxies confirm the presence of this effect (e.g. \cite{Gaztanaga:2002qk,Scranton:2005ci,Menard:2009yb}). 

Many authors have studied the effect of lensing magnification on observations of the galaxy correlation function and power spectrum. Another way of inferring the large-scale mass distribution is through measurements of the neutral Hydrogen density. As light from distant quasars passes through clouds of neutral Hydrogen, photons with rest-frame frequency at the Lyman-$\alpha$ transition ($1216 \AA$) are absorbed. The observed quasar spectrum contains troughs corresponding to absorption by neutral Hydrogen at redshift $z=\nu_\alpha/(\nu_{\rm obs}(1+ v_{\rm pec})) -1$, where $\nu_{\rm obs}$ is the observed frequency of the trough,
$v_{\rm pec}$ is the peculiar velocity at $z$ (speed of light is set to unity), and $\nu_{\alpha}$ is the Lyman-$\alpha$ frequency.\footnote{Here, the redshift $z$ refers to the intrinsic or cosmological redshift, i.e. the redshift if the absorbing materials were comoving. The observed redshift and the intrinsic redshift are related by $1 + z_{\rm obs} = (1 + z)(1 + v_{\rm pec})$.} This Lyman-$\alpha$ forest (of absorption features in the quasar spectrum) is a cosmological tool that can be used to probe the neutral Hydrogen along the line of sight (see for example \cite{Meiksin:2009} and references therein). Here we ask two questions: what is the effect of gravitational lensing on measurements of the Lyman-$\alpha$ forest? And how can we extract the gravitational signal from such measurements?

Let us first discuss qualitatively what we expect to happen. Consider a single line of sight towards a quasar at comoving distance $\chi_Q$. Lensing magnification changes the observed quasar flux $f$, but in a frequency ($\nu$)
independent way 
 \be
\label{Eq:fmag}
  f(\nu)\rightarrow f(\nu)
\cdot \mu (\chi_Q) 
\ee 
where ${\mu}(\chi_Q)$ is the magnification which depends on the density fluctuations along the line of sight. Since all frequencies are treated in the same way, fluctuations in the flux as a function of frequency are \emph{unaffected} --- just as if we were looking at an intrinsically brighter or dimmer quasar. Measurements of for example, the flux power spectrum, from an \emph{individual} line of sight are unchanged by magnification. On the other hand, magnification does change \emph{which} lines of sight are observed. More precisely, a selection bias is introduced: more (fewer) quasars are observed along lines of sight that lead to a positive (negative) magnification correction to quasar number counts. Let ${n}(\chi, \thetaB)$  be the number density of quasars at comoving distance $\chi$ in direction $\thetaB$. Under the effect of lensing magnification,
\be 
\label{nlensed}
{n}(\chi, \thetaB) \rightarrow {n}(\chi, \thetaB) {\mu}(\chi)^{2.5s-1},  \qquad s= {1\over {\rm ln\,} 10} {\int dm \, \epsilon(m) (dn^0/dm) \over \int dm \, \epsilon(m) n^0(m)} \, .
\ee
Here, $n^0(m)$ is the luminosity function of quasars at magnitude $m$ (i.e. $dm \, n^0(m)$ is the number density of quasars at magnitude $m \pm dm/2$), and $\epsilon(m)$ quantifies the sample definition -- the simplest example is a step function which cuts off all quasars fainter than some limiting magnitude $m_{\rm lim.}$, in which case $s$ reduces to the more familiar expression $s = d\log_{10} N(< m_{\rm lim.}) / dm_{\rm lim.}$,  where $N(< m_{\rm lim.})$ is the total number of quasars brighter than $m_{\rm lim.}$. This is a well known result, but a derivation is summarized in Appendix \ref{derive}. With lensing magnification, the lines of sight we observe do not constitute a fair sample of the density field. Therefore, while lensing magnification has \emph{no} effect on measurements of the flux fluctuation from a \emph{single} quasar line of sight, measurements dependent upon the density field \emph{averaged over multiple quasars} \emph{will} be biased.  This occurs since the observable (e.g. the neutral hydrogen density) is correlated with the magnification which depends on the gravitational potential along the line of sight.

What is perhaps the more interesting question is how we could extract the lensing signal from observations of the forest. The discussions above make clear gravitational lensing
affects both the observed brightness and number density of quasars. One could imagine cross-correlating these quantities with the Lyman-alpha forest observables. We will consider several possibilities and identify the ones with an interesting signal-to-noise.

The rest of the paper is organized as follows. In \S\ref{LensBiasDeriv} we develop a simple description of the effect of lensing magnification on measurements of the Lyman-$\alpha$ forest, the key result is Eq.  (\ref{Eq:MagBiasFinal}). In \S \ref{Sims}--\S\ref{FluxPDF} we present the effect of lensing bias on the flux power spectrum, the effective optical depth and the flux probability distribution (PDF) determined by applying a biased weighting scheme to mock absorption spectra from simulations and compare these results with analytical estimates in terms of the flux-mass polyspectra. In \S \ref{MagCorrelation} we discuss the exciting possibility of observing the lensing-induced correlation between quasar magnitudes and the flux fluctuations, the flux power spectrum or the mean flux. While this work is dedicated to discussing magnification bias, dust along the line of sight would have a similar (though generally opposite signed) effect, this is discussed in \S \ref{dust}. We present concluding remarks and discuss the implications of this work for existing and future Lyman-$\alpha$ measurements in \S \ref{Conclusions}. Appendix \ref{derive} contains a unified derivation of the magnitude-forest correlations and the lensing bias discussed in the paper. In Appendix \ref{Extrap} we discuss some issues regarding large-scale flux-mass correlations measured from our simulations and the accuracy of an analytic description. Appendix \ref{estimator} derives an estimator for the flux-magnitude correlation and its associated error.  

Before proceeding we should mention some related literature. The magnification bias to the statistics of metal absorption lines in quasar spectra was investigated in \cite{Thomas:1990aj} and a method for detecting statistical lensing by absorbers was proposed in \cite{Menard:2005}. Lensing effects on the statistics of damped Lyman-$\alpha$ systems and the inferred density of neutral Hydrogen was considered in \cite{Bartelmann:1996}. More recently, \cite{Menard:2003vf} studied magnification bias due to intervening absorbers in the 2dF quasar survey and \cite{Menard:2008} in SDSS. Recent work by \cite{Vallinotto:2009wa,Vallinotto:2009jx} proposed correlating lensing in the cosmic microwave background with fluctuations in the forest to extract the flux-mass correlation.

\section{Lensing as Noise/Bias}
\label{lensingnoise}

\subsection{Formalism}
\label{LensBiasDeriv}

We are interested in some Lyman-alpha forest observable $\justO$, which can represent the flux power spectrum, the flux transmission/decrement,
the flux fluctuation (around its mean), and so on. Let us use $O_I$ to denote the observable measured from a quasar labeled by $I$. We typically form an estimator by averaging over quasars:
\be 
\label{Eq:Osum}
{\mathcal{O}}_{\rm obs}=\frac{\sum_I w_I {\mathcal{O}}_I}{\sum_I w_I} 
\ee 
where $w_I$ denotes weights, the simplest example of which is $w_I = 1$.

Two important points. First $\justO_I$, the observable on a quasar by quasar basis, is generally {\it not} affected by gravitational lensing. This is because gravitational lensing affects all wavelengths equally. For instance, $\justO_I$ could represent the flux transmission which is $f/f_C$ (where $f$ and $f_C$ are the flux and continuum respectively as a function of frequency), or the flux fluctuation $\delta_f = (f - \bar f)/\bar f$ (where $\bar f$ is the flux averaged for the particular line of sight in question). Since gravitational lensing brightens or dims $f$, $\bar f$ and $f_C$ all equally independent of wavelength, there is no effect on $\justO_I$.

The other important point is that $\justO_{\rm obs}$ {\it is} generally affected by  lensing. The crucial observation is that $w_I$ in Eq. (\ref{Eq:Osum}) reveals only part of the weighting that is going on; in any given dataset, we inevitably give zero weights to quasars which are too faint to observe. This means that even if one chooses $w_I = 1$, one is merely performing a straight average over one's sample, as opposed to an average over all possible quasars. To account for this fact, it is helpful to  pixelize the sky, and rewrite ${\justO}_{\rm obs}$ as 
\begin{eqnarray}
\label{Eq:Osum2}
{\justO}_{\rm obs} = {\sum_i w_i n_i \justO_i \over \sum_i w_i n_i}
\end{eqnarray}
where $i$ is the pixel label. One could conceptually think of each pixel as sufficiently small that the number of quasars in it $n_i$ is either $1$ or $0$; we will more generally think of $n_i$ as simply the number density of quasars in pixel $i$. This way of rewriting is useful because it makes explicit the fact that $\justO_{\rm obs}$ is a weighted average -- it is weighted by the number density of quasars, on top of whatever additional weighting $w_i$ one might wish to apply. In other words, since our Lyman-alpha forest observable can be measured only if there exists a background quasar in the sky location of interest, the observable is always implicitly weighted by the abundance of quasars.

Eq. (\ref{nlensed}) tells us that
\begin{eqnarray} 
n_i = n_i^{\rm intrinsic} (1 + \delta^\mu_i)
\end{eqnarray}
where $n_i^{\rm intrinsic}$ is the intrinsic pre-lensed quasar number density, and $\delta^\mu_i$ is the lensing modification given by
\begin{eqnarray}
\label{Eq:KappaDef}
\delta^\mu_i \equiv (5s - 2) \kappa_i \quad , \quad {\kappa}_i =\int_0^{\chi_Q} d\chi'\frac{\chi_Q-\chi'}{\chi_Q}\chi' \nabla_\perp^2\phi(\chi', i)
\end{eqnarray}
assuming fluctuations are small.  The expression for the convergence $\kappa_i$ assumes spatial flatness, but can be easily generalized to non-flat universes ($\chi_Q$ is the comoving distance to quasar).  Here, $1 + 2\kappa$ is the weak lensing approximation to the lensing magnification $\mu$ of Eqs. (\ref{Eq:fmag}) and (\ref{nlensed}) (the full nonlinear expression is given in footnote \ref{nonlinfootnote}). Using $w_i = 1$ in Eq. (\ref{Eq:Osum2}), assuming there is no correlation between the forest observable and the {\rm intrinsic} quasar number fluctuation, and ignoring corrections of the integral constraint type (see Appendix \ref{derive} and for example, \cite{Hui:1998ix}), we obtain the following ensemble average,  to the lowest order in fluctuations:
\begin{eqnarray}
\langle \justO_{\rm obs} \rangle = \justO_{\rm true} + \langle \justO_i \delta^\mu_i \rangle
\end{eqnarray}
where $\langle \justO_i \delta^\mu_i \rangle$  is the correlation between the observable and the magnification fluctuation at the same $i$ (i.e. zero lag). Dropping the $i$ label, the lensing induced measurement bias on an observable $\justO$ is therefore
\begin{eqnarray}
\label{Eq:MagBiasFinal}
\langle \justO_{\rm obs} \rangle - \justO_{\rm true} = \langle \justO \delta^\mu \rangle = (5s-2) \langle \justO \kappa \rangle \, .
\end{eqnarray}
This is a fundamental expression that we will use repeatedly below. The meaning of each of these symbols is as follows: $\justO$ is the fluctuation which we attempt to measure with the estimator $\justO_{\rm obs}$, which upon ensemble averaging  generally differs from the true underlying value $\justO_{\rm true} = \langle \justO \rangle$.

It is important to emphasize the actual measurement bias could be different. The above estimate assumes that all quasars {\it within one's sample} are weighted equally (i.e. $w_i = 1$).  In practice, one might want to weigh brighter quasars {\it within one's sample} more strongly. For instance, one could weigh by the net flux of the quasar, which corresponds to inverse variance weighting in the noise dominated regime. This would result in a lensing induced measurement bias of (see Appendix \ref{derive} for derivation)
\begin{eqnarray}
\label{Eq:MagBiasFluxWeight}
\langle \justO_{\rm obs} \rangle - \justO_{\rm true} = (5s' - 2) \langle \justO \kappa \rangle \, .
\end{eqnarray}
where $s'$ is defined as
\begin{eqnarray}
\label{sprime}
s' = {1\over {\rm ln\,} 10} {\int dm \, \epsilon(m) (dn^0/dm) 10^{-m/2.5} \over\int dm \, \epsilon(m) n^0(m) 10^{-m/2.5}}
\end{eqnarray}
which can be contrasted with $s$ defined in Eq. (\ref{nlensed}). For a step function sample definition $\epsilon$, $s'$ reduces to $d\log_{10} N'(< m_{\rm lim.}) / dm_{\rm lim.} + 0.4$, where $N' = \int_{-\infty}^{m_{\rm lim.}} dm \, n^0 10^{-m/2.5}$. This is generally larger than $s$.

The precise value of $s$ or $s'$ is sample dependent. The relatively low redshift ($1<z<2.2$) Sloan Digital Sky Survey (SDSS)  quasars that were used to measure the magnification-galaxy cross-correlation  have a slope that spans $-1 \lsim 5s-2 \lsim 1.9$, depending on  the magnitude cut \cite{Scranton:2005ci}. For this paper, the higher redshift quasars for which the Lyman-alpha forest falls within the SDSS spectral range are more relevant. In Fig. \ref{Fig:slope} (in Appendix \ref{derive}) we show the cumulated number counts and rough estimates of $s$ and $s'$ from SDSS data release $6$.

In the rest of this paper, we will adopt $s = 1$ (or $s' = 1$), so our results for the lensing biases can be scaled up and down by $(5s-2)/3$ or $(5s'-2)/3$.

We will also adopt the following set of cosmological parameters in presenting numerical estimates. We assume a flat $\Lambda$CDM cosmology with cosmological parameter values $\Omega_m=0.3$, $\Omega_\Lambda=0.7$, $\Omega_b=0.04$ for the fractional densities in matter, vacuum and baryons;  scalar spectral index $n_s=1$; Hubble parameter today $H_0=100h$ km/s/Mpc with $h=0.7$ and fluctuation amplitude $\sigma_8=0.9$. The speed of light is set to unity. In a few places we use analytic calculations of the matter and baryon power spectra. The 3-D matter power spectrum is calculated using the transfer function of \cite{Bardeen:1985tr} with a modified shape parameter $\Gamma=\Omega_mh {\,\rm exp}\,[{-\Omega_b(1+\sqrt{2h}/\Omega_m)}]$ \cite{Sugiyama:1994ed}, and the non-linear evolution is modeled according to \cite{Peacock:1996ci}. To relate baryons to mass we assume the 3-D Fourier space fluctuations are related by $\delta_b(\vk)={\,\rm exp}\,[{-k^2/k_s^2}]\delta (\vk)$, where $k_s= k_J \sqrt{10/3}$ and $k_J$ is the Jeans scale (see Appendix \ref{Extrap} for more details).

\subsection{Lensing Bias from Simulations}
\label{Sims}

Lensing magnification biases measurements of the forest so long as the forest observable is correlated with the lensing convergence (Eq. (\ref{Eq:MagBiasFinal})). Since features in the Lyman-$\alpha$ forest are caused by absorption by gas which itself traces gravitational potentials, there ought to be a correlation between absorption features and the lensing convergence. However, the mapping between absorption-induced fluctuations in the flux and the gravitational potential is nonlinear and dependent on potentially uncertain physics, such as fluctuations in the ionizing background. For this reason we need to appeal to simulations to quantify the correlation between forest observables (e.g. $\delta_f$ and $P_{ff}$) and the lensing convergence. 

In this section we use hydrodynamic simulations to study the effect of magnification bias on the Lyman-$\alpha$ forest. Specifically, we study the ``D5" simulation of \cite{Springel:2002uv}. This simulation was run using an entropy-conserving \cite{Springel:2001qb} version of the smoothed particle hydrodynamics (SPH) code Gadget \cite{Springel:2000yr}. It tracks $324^3$ dark matter particles and $324^3$ gas particles in a simulation volume of co-moving side length $L = 33.75$ Mpc/$h$, and includes a subresolution model for star formation \cite{Springel:2002ux} and heating by a uniform background radiation field \cite{Katz:1995up}. Further details regarding the simulation can be found in \cite{Springel:2002uv}. We generate mock Lyman-alpha forest spectra along random lines of sight through the simulation volume in the usual way, integrating through the SPH kernels of the particles, and incorporating the effect of peculiar velocities and thermal broadening (e.g. \cite{Hernquist:1995uma}). We generate mock spectra from simulation snapshots at $z=2,3$ and $4$. In each case, we adjust the amplitude of the photoionizing background in the simulation to match the observed mean transmitted flux from \cite{FaucherGiguere:2007ys}. We also make use of the ``G6" simulation (see \cite{Nagamine:2005jp}), which tracks $2 \times 486^3$ particles in a $L = 100$ Mpc/$h$ box, in order to check the sensitivity of our results to finite boxsize effects in Appendix \ref{Extrap}. We have checked that the flux power spectra from these simulations approximately match the existing measurements.

Magnification bias biases which lines of sight an observer will see: sight lines with positive fluctuations in magnification are more abundant and those with negative less abundant. With simulations we can of course see every line of sight (a mock spectrum is generated regardless of whether there is a quasar behind it) so we mimic lensing bias by weighting each line-of-sight $\thetaB$ by  $1+\delta^\mu(\thetaB)$.
The simulations are in boxes of size $33.75 \MpcOh$, so we can't actually calculate the lensing convergence $\kappa$ along the entire line-of-sight but instead for the weighting of absorption measurements at redshift $\bar{z}$ we use\footnote{While it would be more accurate to leave the $\chi$ terms inside the integral, their amplitude is slowly varying across the length of our simulations box so Eq. (\ref{eq:weightdef}) shouldn't be a bad approximation.}
\be
\label{eq:weightdef}
w_{\thetaB}=1+\delta^\mu (\thetaB) \sim
1+\frac{3}{2}H_0^2\Omega_m(5s-2)\frac{\chi_Q-\bar{\chi}}{\chi_Q}\bar{\chi}(1+\bar{z})\int_{\bar{\chi}}^{\bar{\chi}+L} d\chi' \delta(\chi', \thetaB)
\ee
where $L$ is the length of the box $\bar{\chi}=\chi(\bar{z})$, $\chi_Q$ is the distance to the quasar, $\delta$ is the mass fluctuation, and we have used the approximation $\nabla_\perp^2\phi\approx\nabla^2\phi=\frac{3}{2}H_0^2\Omega_m(1+z)\delta$ which should be valid under the line-of-sight integral. Since the bias only comes from the mass
fluctuations $\delta$ which are correlated with flux fluctuations $\delta_f$ at the redshift we are considering, and correlations drop off rapidly with increased spatial separation, neglecting contributions to $\kappa$ from mass fluctuations at lower redshifts shouldn't  be a bad approximation. We will verify this below by comparing the lensing bias thus obtained with that obtained from another method.

\subsection{Bias to The Flux Power Spectrum}
\begin{figure}
\begin{center}
$\begin{array}{cc}
\includegraphics[width=0.45\textwidth]{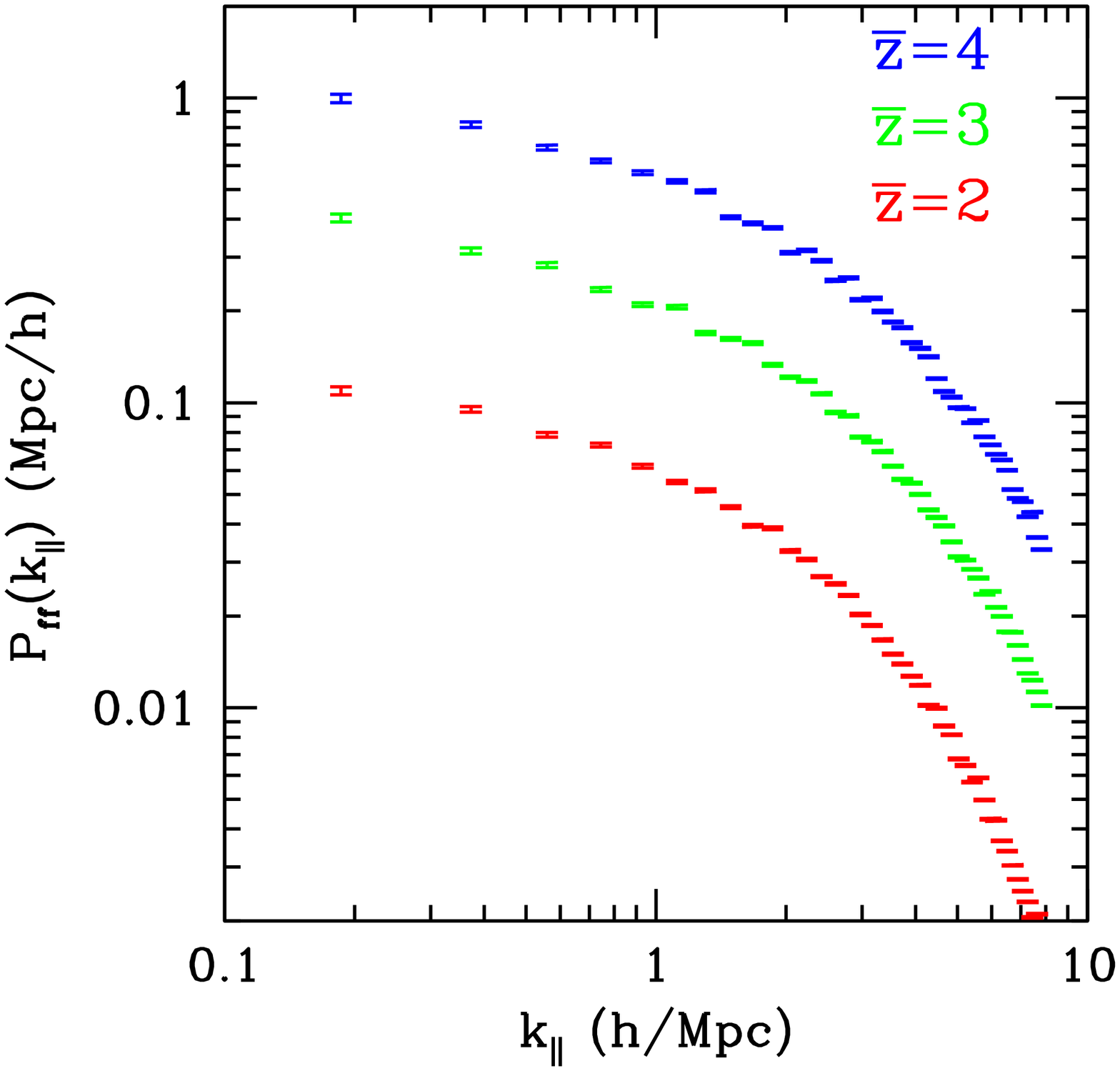}&\includegraphics[width=0.45\textwidth]{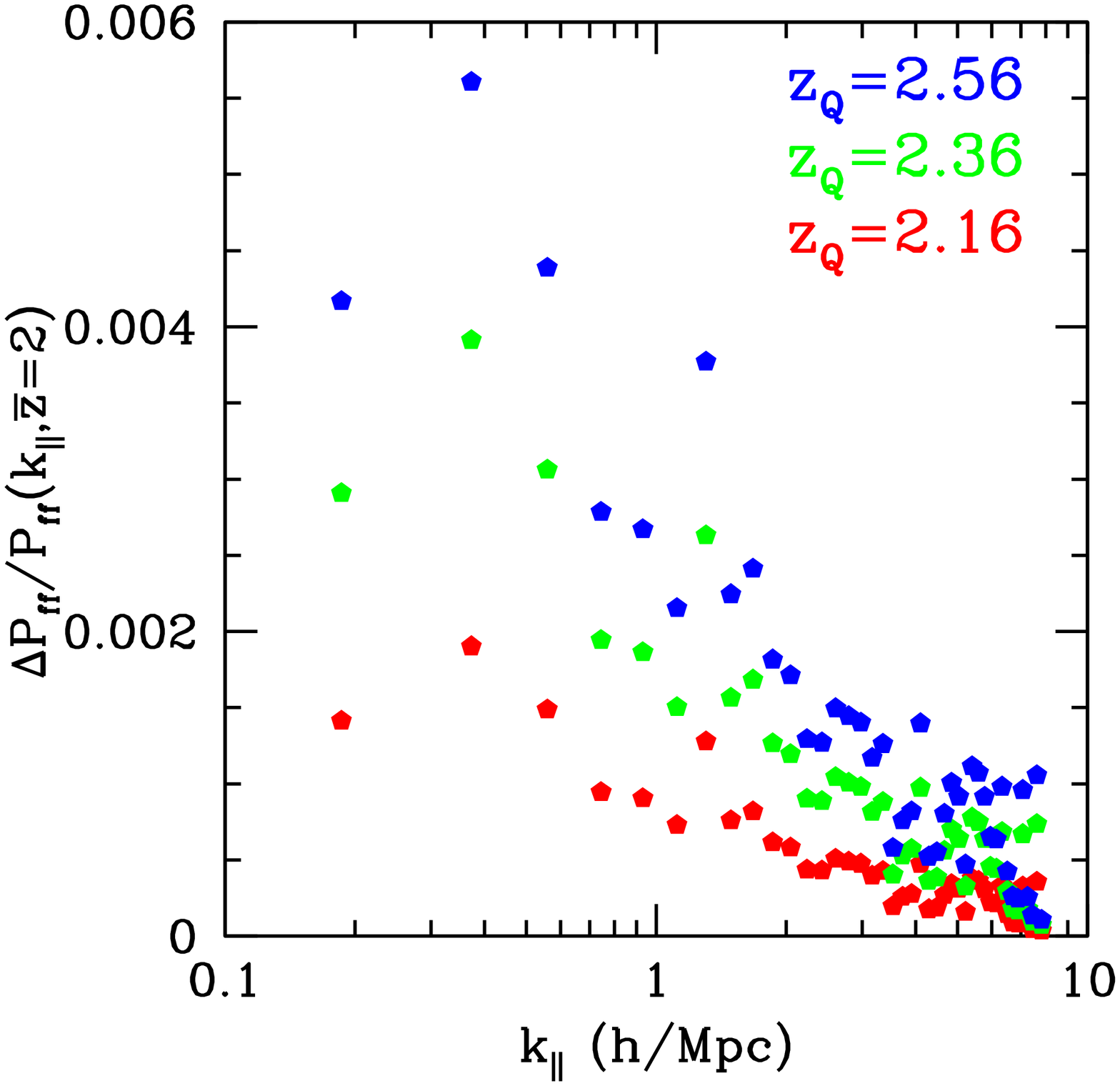}\\
\mbox{(a)}&\mbox{(b)}\\
\includegraphics[width=0.45\textwidth]{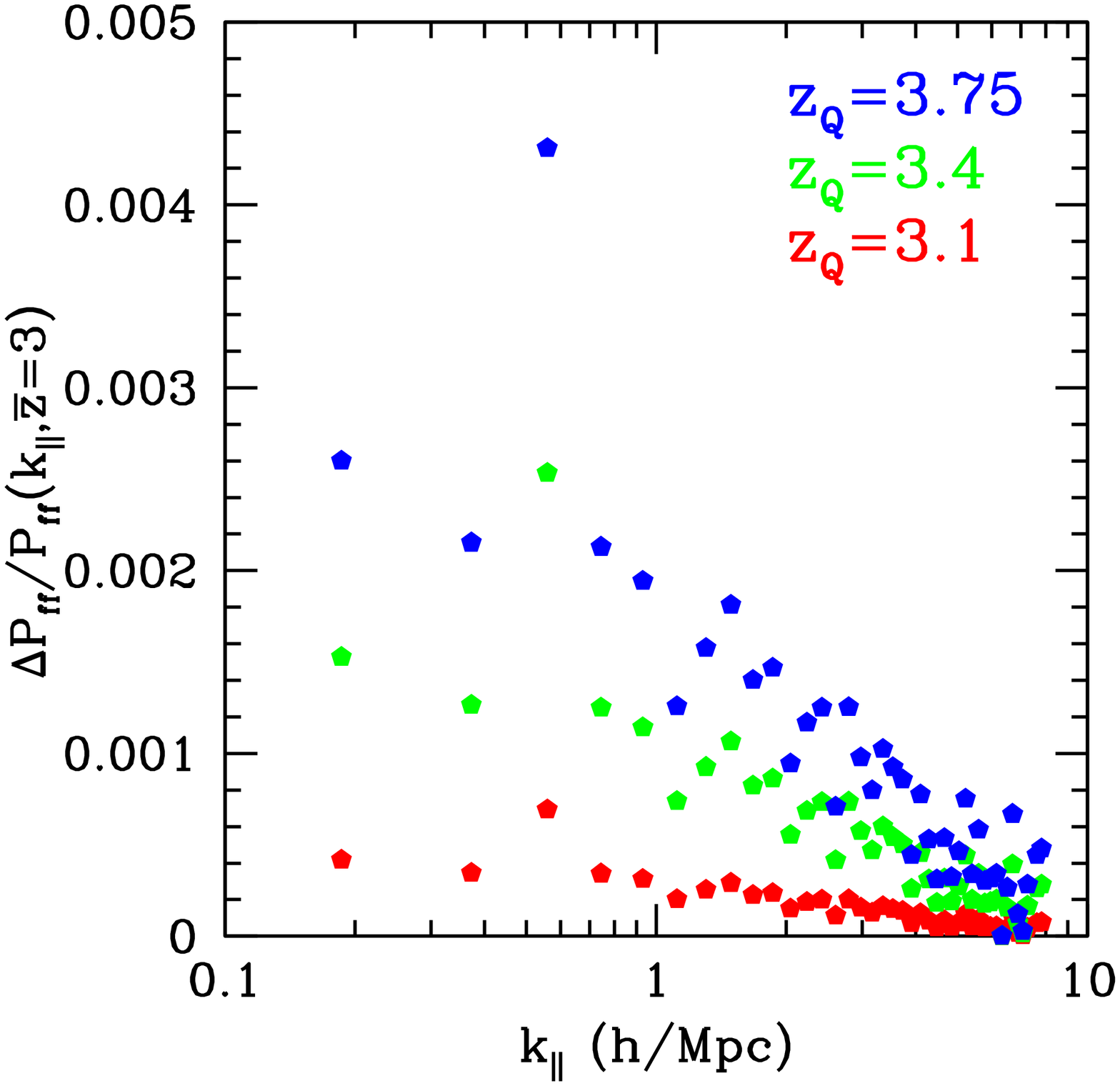}&\includegraphics[width=0.45\textwidth]{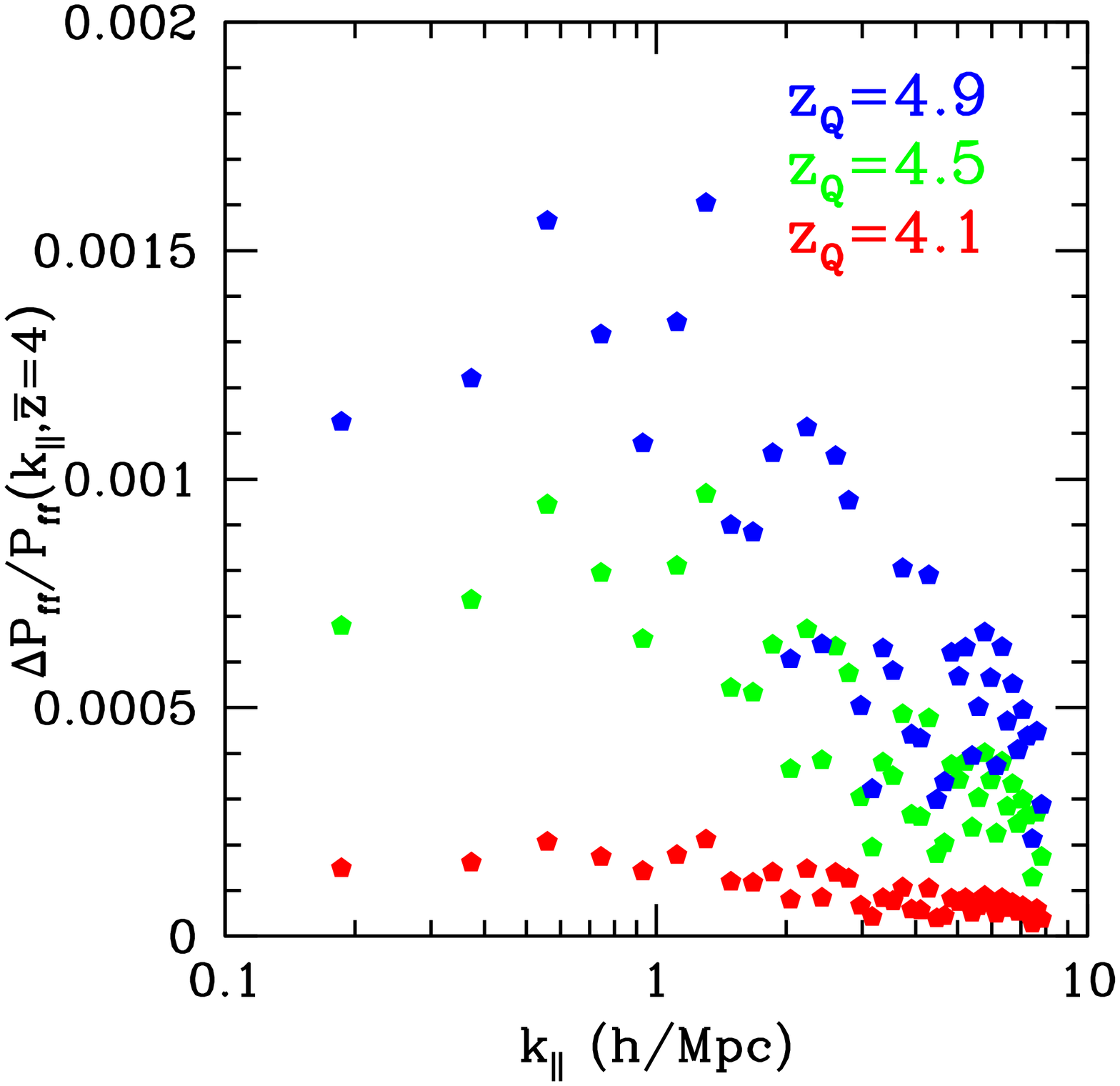}\\
\mbox{(c)}&\mbox{(d)}
\end{array}$
\caption{\label{Fig:PffSims}(a) The 1D flux power spectrum measured from the simulations. Panels $(b)$ - $(d)$ show the fractional correction from magnification 
$(\langle P_{ff} {}_{\rm obs} \rangle - P_{ff} {}_{\rm true})/P_{ff} {}_{\rm true}$
for mean redshifts $\bar{z}=2, 3, 4$ respectively. The error bars are suppressed for clarity. The amplitude of the correction depends on the redshift of the quasar $z_Q$ through $\kappa$ (Eq. (\ref{Eq:KappaDef})) and also on the slope of the quasar number count function, here we assume $5s-2=3$.}
\end{center}
\end{figure}

In panel $(a)$ of Fig. \ref{Fig:PffSims} we show the 1D flux power spectrum
\be
P_{ff}(k_{\parallel})=\frac{1}{L}|\delta(k_{\parallel})|^2
\ee
measured from simulations\footnote{Our Fourier convention is $\delta(\vk)=\int \!d^3\vx \,e^{-i\vk\cdot\vx}\delta(\vx)$.}. In panels $(b)$-$(d)$ we show the fractional correction due to magnification, calculated by weighting lines-of-sight by $1+\delta_\mu$.  Magnification weighs lines of sight with large $\kappa$ more heavily than those with low $\kappa$. Measurements of $\delta_{f}$ at redshift $\bar{z}$ can be taken from quasars at redshift $z_Q$ for $\bar{z} \lsim z_Q \lsim (1+\bar{z})\nu_\beta/\nu_\alpha-1$. The upper limit is set by where the Lyman-$\beta$ line ($\lambda_\beta=1026\AA$)  can be confused with Lyman-$\alpha$. The lensing weight function is peaked at $\chi\sim \chi_Q/2$, so for quasars at higher redshift, $\chi(\bar{z})$ is closer to the peak of the lensing weight function making the lensing correction larger than it is for quasars just beyond $\bar{z}$. 

Magnification causes a $\lsim 1\%$ change to $P_{ff}$ with larger changes occurring at lower redshifts.   The results shown in Fig. \ref{Fig:PffSims} assume $s = 1$ and that all quasars within one's sample are weighed equally. The actual lensing correction in realistic surveys could differ by factor of a few. The current statistical error on the amplitude of $P_{ff}$, averaged over scales, is $\sim 0.6\%$ from SDSS \cite{McDonald:2004eu}, which is a bit larger than the magnification correction shown in Fig. \ref{Fig:PffSims}.
However, as precision improves, a systematic offset such as that introduced by lensing could well be relevant.

\label{PCorrection}

It is instructive to get an analytic estimate for the bias to the flux power spectrum.  Equation (\ref{Eq:MagBiasFinal}) tells us the lensing bias for the flux power spectrum is:
\begin{eqnarray}
\label{Pffbias}
\langle P_{ff} {}_{\rm obs} (k_\parallel) \rangle - P_{ff} {}_{\rm true} (k_\parallel)= {1\over \Delta\chi}  \langle \delta_f (k_\parallel) \delta_f^* (k_\parallel) \delta^\mu \rangle
\end{eqnarray}
where $\Delta\chi$ is the length of the quasar spectra from which the power is measured. It is therefore clear that this lensing bias depends on the three-point function  or bispectrum. Expressing $\delta^\mu$ as a line-of-sight integral of the mass fluctuation (Eq. (\ref{eq:weightdef})), and  judiciously applying the Limber's approximation \cite{Limber:1954,LoVerde:2008re}, we obtain:
\begin{eqnarray}
\label{Eq:DPBffmHighK}
\langle P_{ff} {}_{\rm obs} (k_\parallel) \rangle - P_{ff} {}_{\rm true} (k_\parallel) \approx \frac{3}{2}H_0^2\Omega_m(5s-2) \frac{\chi_Q-\bar\chi}{\chi_Q}\bar\chi (1+\bar z) B_{ffm}(k_{||},-k_{||},0)\, .
\end{eqnarray}
where $\chi_Q$ is the distance to quasar, $\bar\chi$ is the average distance  to the forest, and $\bar z$ is the corresponding redshift. Here, $B_{ffm}$ is the (true) 1D flux-flux-mass bispectrum at redshift $\bar z$. It should be emphasized that $B_{ffm} (k_\parallel, -k_\parallel, 0)$  does not vanish despite having one of its arguments (the one associated with mass) being zero. This is related to the fact that it arises from a 1D projection of a 3D distribution. For instance, it is well known that a 3D power spectrum and its 1D projection are related by $P_{1D} (k_\parallel) = \int_{k_\parallel}^\infty (k dk /2\pi) P_{3D} (k)$, and so $P_{1D}$ generally does not vanish in the $k_\parallel \rightarrow 0$ limit.

We calculate the correction to $P_{ff}$ using $B_{ffm}$ measured from simulations.  Estimating $B_{ffm} (k_\parallel, -k_\parallel, 0)$ requires an extrapolation from what we actually measure $B_{ffm} (k_\parallel, -(k_\parallel + \Delta k), \Delta k)$, where the smallest $\Delta k$ is the fundamental mode of the box. How this is done is described in detail in Appendix \ref{Extrap}. In Fig. \ref{Fig:PffBispec} we compare the lensing correction to the flux power spectrum  as determined via the bispectrum to that determined by the weighting method described previously in \S \ref{Sims}. The qualitative agreement between the two is reassuring.

\subsection{Bias to The Effective Optical Depth}
\label{SimsTau}
\begin{figure}
\begin{center}
\includegraphics[width=0.45\textwidth]{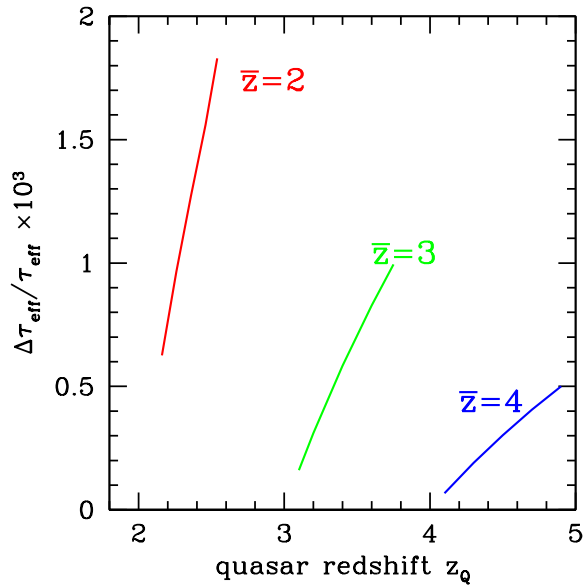}
\caption{\label{Fig:TauBias} The magnification correction to the effective optical depth at three different mean redshifts $\bar{z}$. This correction depends on the redshift $z_Q$ of the quasars. In practice, measurements of $\tau_{\rm eff}$ involves combining
source quasars at multiple redshifts. In the above plot we assume  $5s-2=3$.}
\end{center}
\end{figure}

The effective optical depth $\tau_{\rm eff}$  is estimated from data by averaging over frequencies (and quasars) the ratio of the observed flux $f$= to the continuum $f_C$, i.e. $e^{-\tau_{\rm eff}}$  is estimated from $f/f_C$. As emphasized in \S \ref{LensBiasDeriv}, lensing magnification affects $f$ and $f_C$ in the same manner, and therefore leaves $f/f_C$ unchanged on a quasar by quasar basis. Lensing's impact enters through the sample selection, resulting in (see Eq. (\ref{Eq:MagBiasFinal}))
\begin{eqnarray}
\label{taueffbias}
\langle e^{-\tau_{\rm eff}} {}_{\rm obs} \rangle - e^{-\tau_{\rm eff}}{}_{\rm true} = \langle e^{-\tau_{\rm eff}} \delta^\mu \rangle
\end{eqnarray}
where $e^{-\tau_{\rm eff}}$ on the right is the mean transmission on a line-of-sight by line-of-sight basis, and $e^{-\tau_{\rm eff}} {}_{\rm true}$ is the true mean transmission if only we could  dispense with the quasar-weighting and average over the whole sky equally.

The correction to the effective optical depth measured from simulations is shown in Fig. \ref{Fig:TauBias} for quasars at several redshifts.  The net effect of lensing will depend on how measurements from quasars at different redshifts are combined. However, Fig. \ref{Fig:TauBias} indicates the effect should be $\sim 10^{-3}-10^{-4}$. Current measurements of $\tau_{\rm eff}$ have errors of at least a few percent at each redshift so at present lensing should not be an issue.\footnote{\label{continuumfootnote} The error bars for measurements of $\tau_{\rm eff}$ are dominated by uncertainties in the continuum-fit. Typically, the continuum is estimated either by extrapolating from the red side
(e.g. \cite{Bernardi:2002mr}) , or by performing a smooth fit through portions of the spectra that are deemed unabsorbed (e.g. \cite{FaucherGiguere:2007ys}). We simulate the effect of the latter by renormalizing the flux along each line of sight by $f_{\rm max}$, the maximum along that line of sight. At low redshifts, this introduces a
negligible bias to $\tau_{\rm eff}$, but by $z = 4$, $\tau_{\rm eff}$ is biased low by $\sim 8 \%$, in rough agreement with \cite{FaucherGiguere:2007ys}.
However, we have checked that the fractional magnification correction is not significantly changed whether we renormalize by $f_{\rm max}$ or not.}

\label{TauCorrection}

An analytical estimate of the lensing bias on the effective optical depth can be inferred from Eq. (\ref{taueffbias}), which gives: 
\ba
\label{Eq:TauBias}
{\langle e^{-\tau_{\rm eff}} {}_{\rm obs} \rangle \over e^{-\tau_{\rm eff}} {}_{\rm true}}- 1 = \langle \delta_f \delta^\mu \rangle \approx(5s-2)\frac{3}{2}\Omega_mH_0^2\frac{\chi_Q-\bar\chi}{\chi_Q}\bar\chi(1+\bar z)P_{fm}(k_{||}=0)
\ea
where we have used Limber's approximation, $P_{fm}(k_{||})$ is the (true) 1D flux-mass  power spectrum evaluated at the mean redshift of absorption $\bar z$, $\bar\chi$ is
the corresponding distance, and $\chi_Q$ is the quasar distance.  The lensing bias of this one-point statistic is thus related to a two-point correlation, just as the lensing bias of the power spectrum is determined by the bispectrum. As before, to evaluate the bias using this method,  an extrapolation of the simulation $P_{fm}$ to $k_\parallel = 0$ is necessary (see Appendix \ref{Extrap}).  We find results that are consistent with those obtained by the weighting method (Fig. \ref{Fig:TauBias}) to about $30 \%$,
with the latter generally higher.

\subsection{Bias to The Flux Probability Distribution Function}

\begin{figure}
\begin{center}
$\begin{array}{cc}
\includegraphics[width=0.45\textwidth]{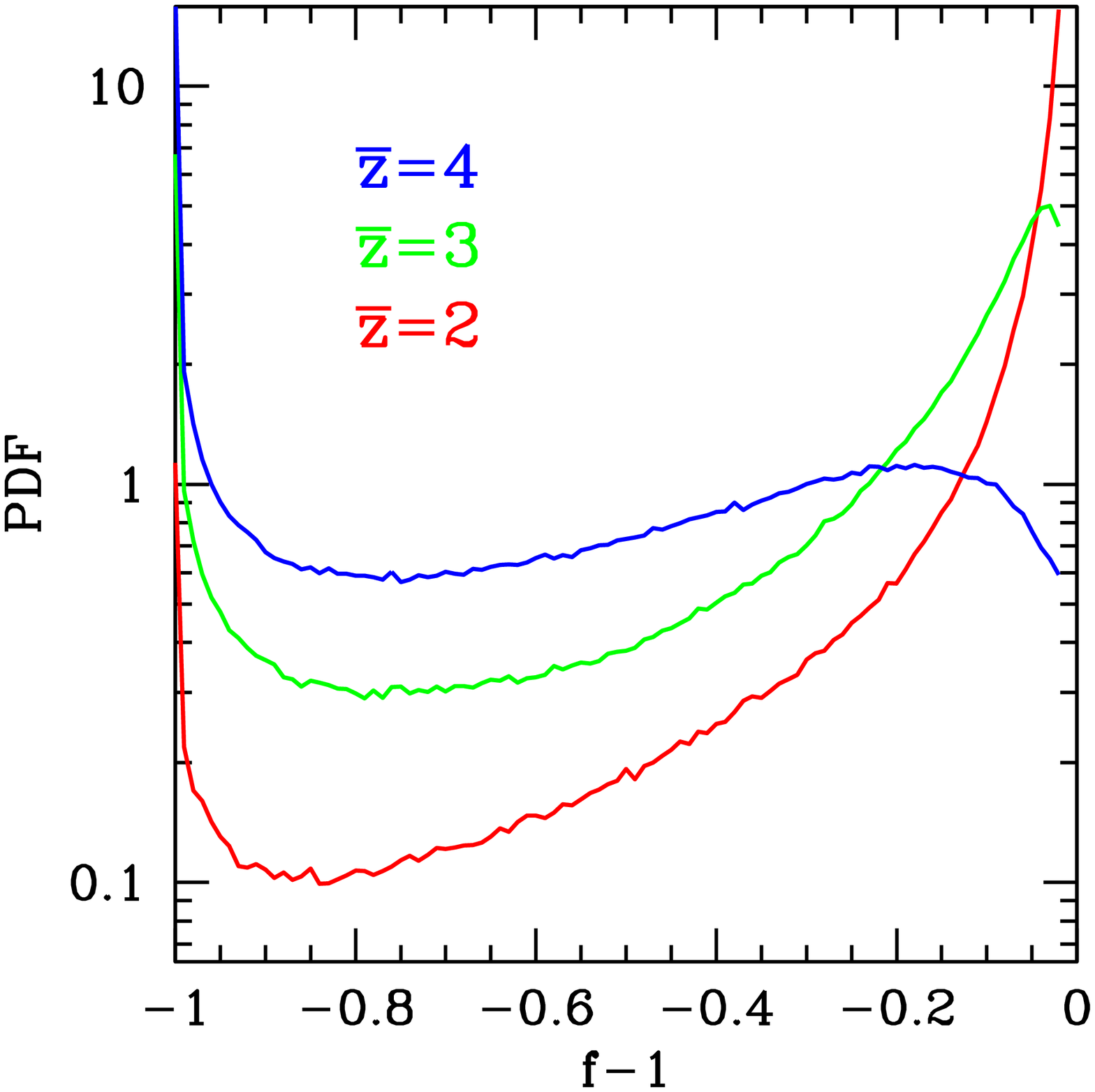}&\includegraphics[width=0.45\textwidth]{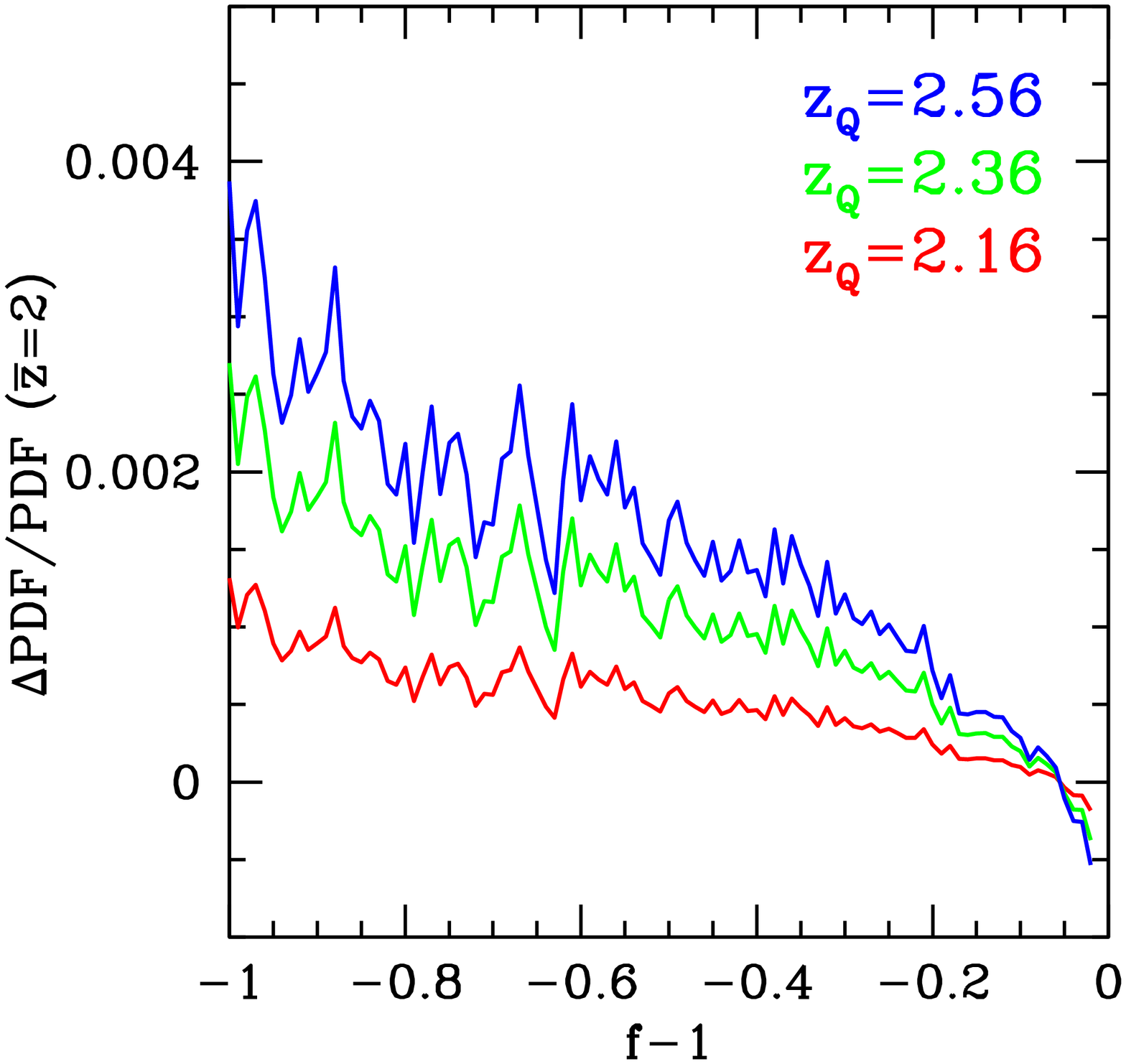}\\
\mbox{(a)}&\mbox{(b)}\\
\includegraphics[width=0.45\textwidth]{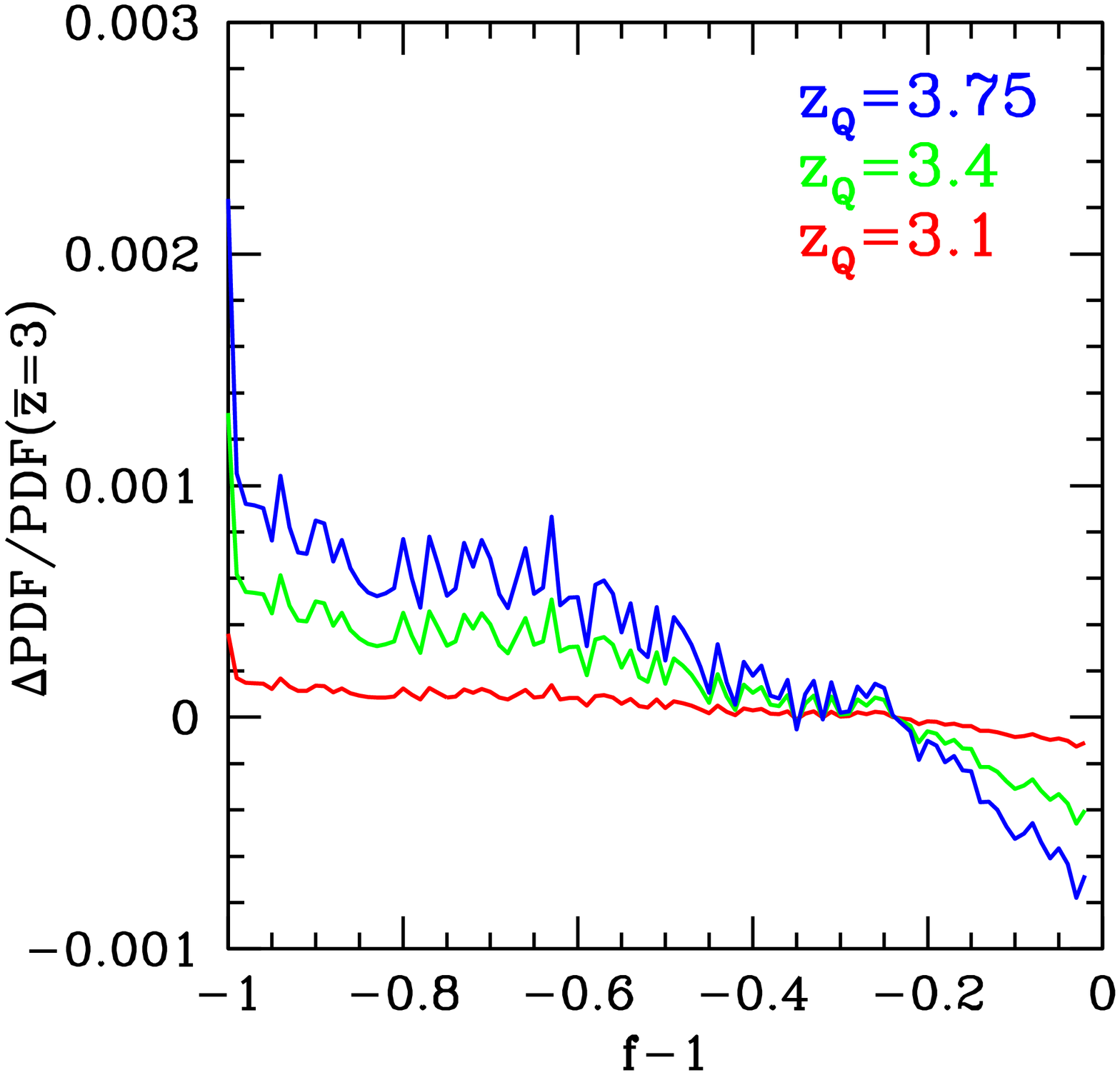}&\includegraphics[width=0.45\textwidth]{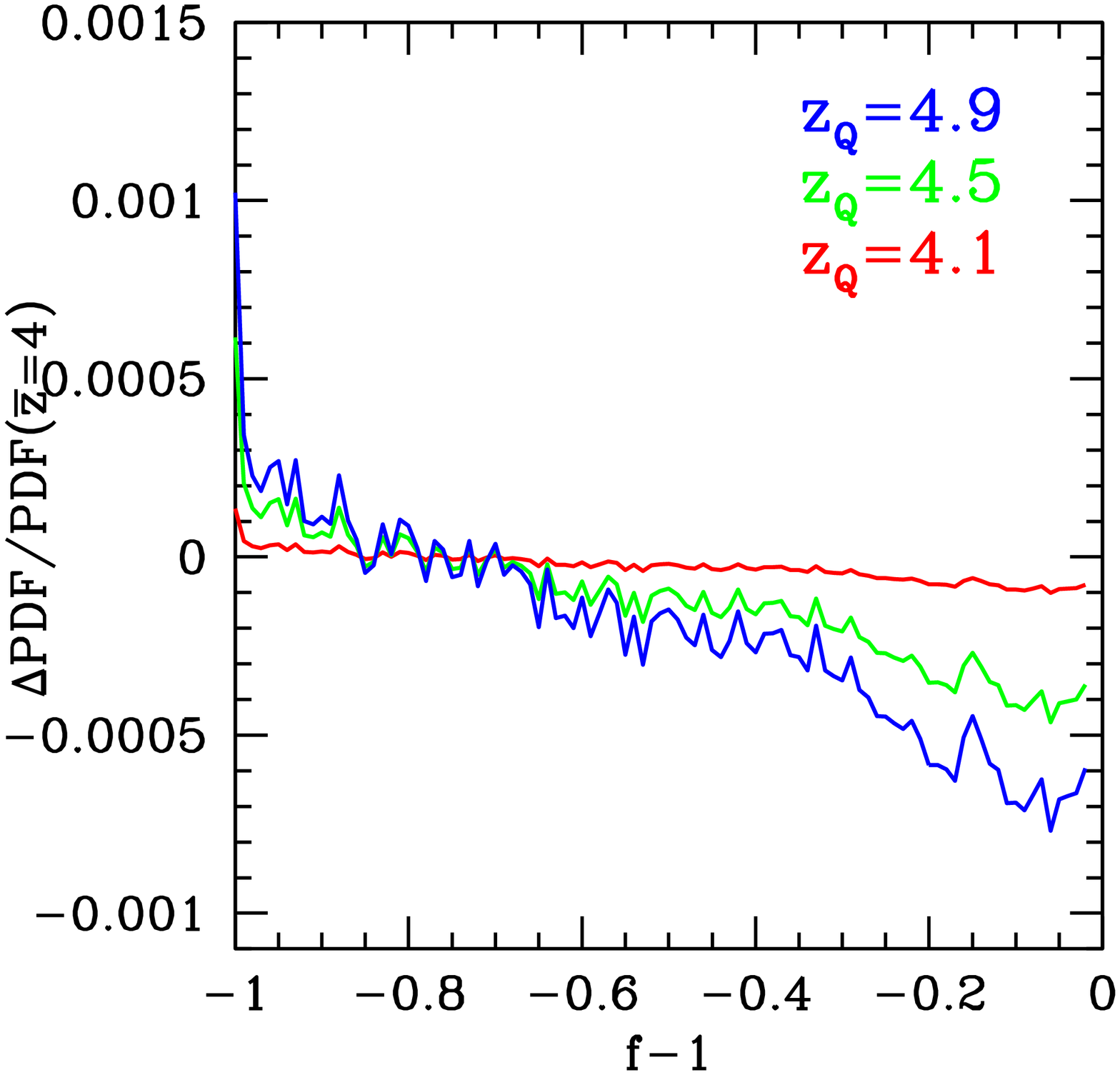}\\
\mbox{(c)}&\mbox{(d)}
\end{array}$
\caption{\label{Fig:PDFetauSims}(a) The flux probability distribution as measured from the simulations. Panels $(b)$ - $(d)$ show the fractional correction from magnification for mean redshifts $\bar{z}=2, 3, 4$. The amplitude of the correction depends on the redshift of the quasar $z_Q$ and the slope of the quasar number count function, here we assume $5s-2=3$ (Eq. (\ref{Eq:KappaDef})).}
\end{center}
\end{figure}

\begin{figure}
\begin{center}
$\begin{array}{cc}
\includegraphics[width=0.45\textwidth]{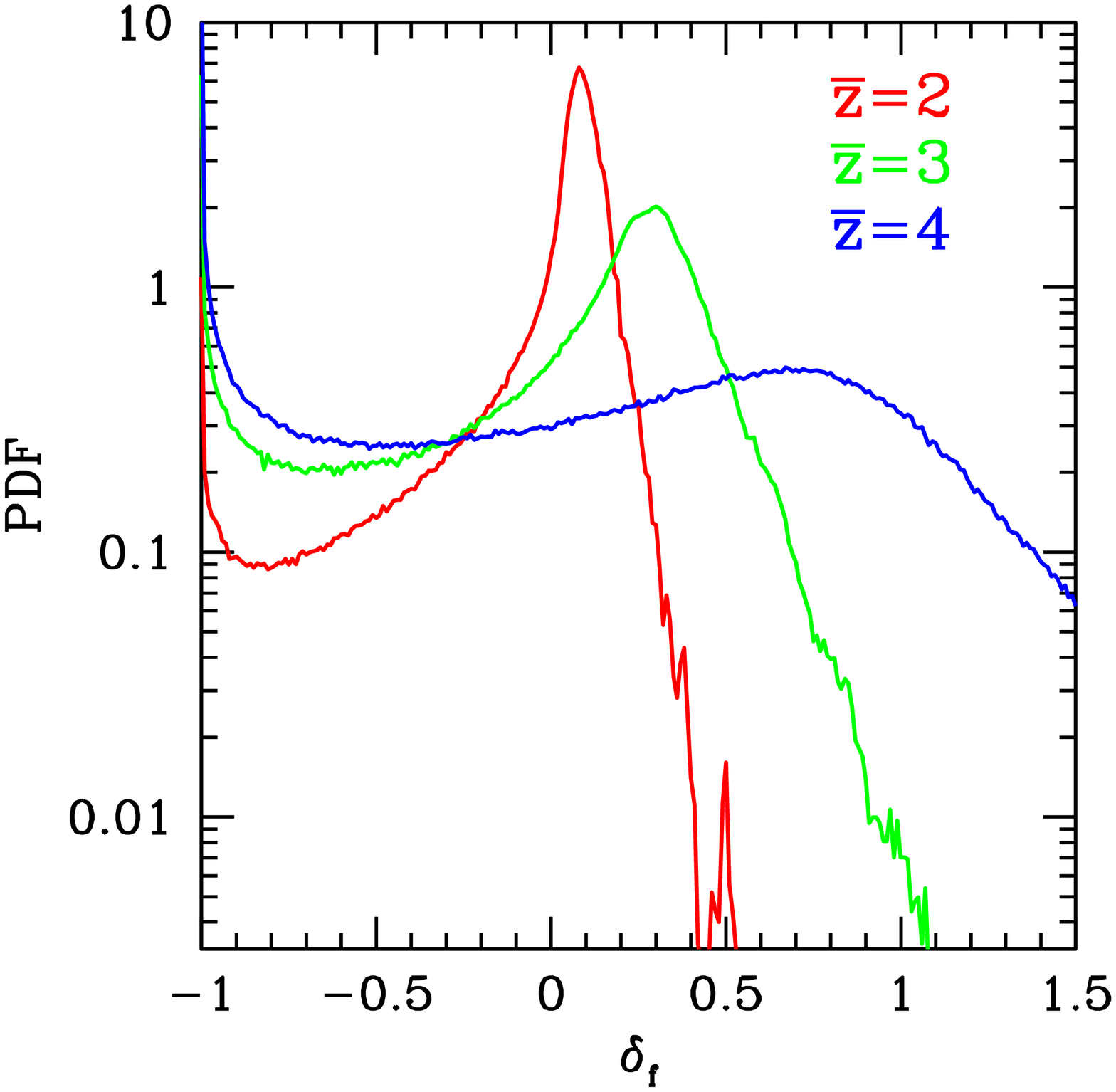}&\includegraphics[width=0.45\textwidth]{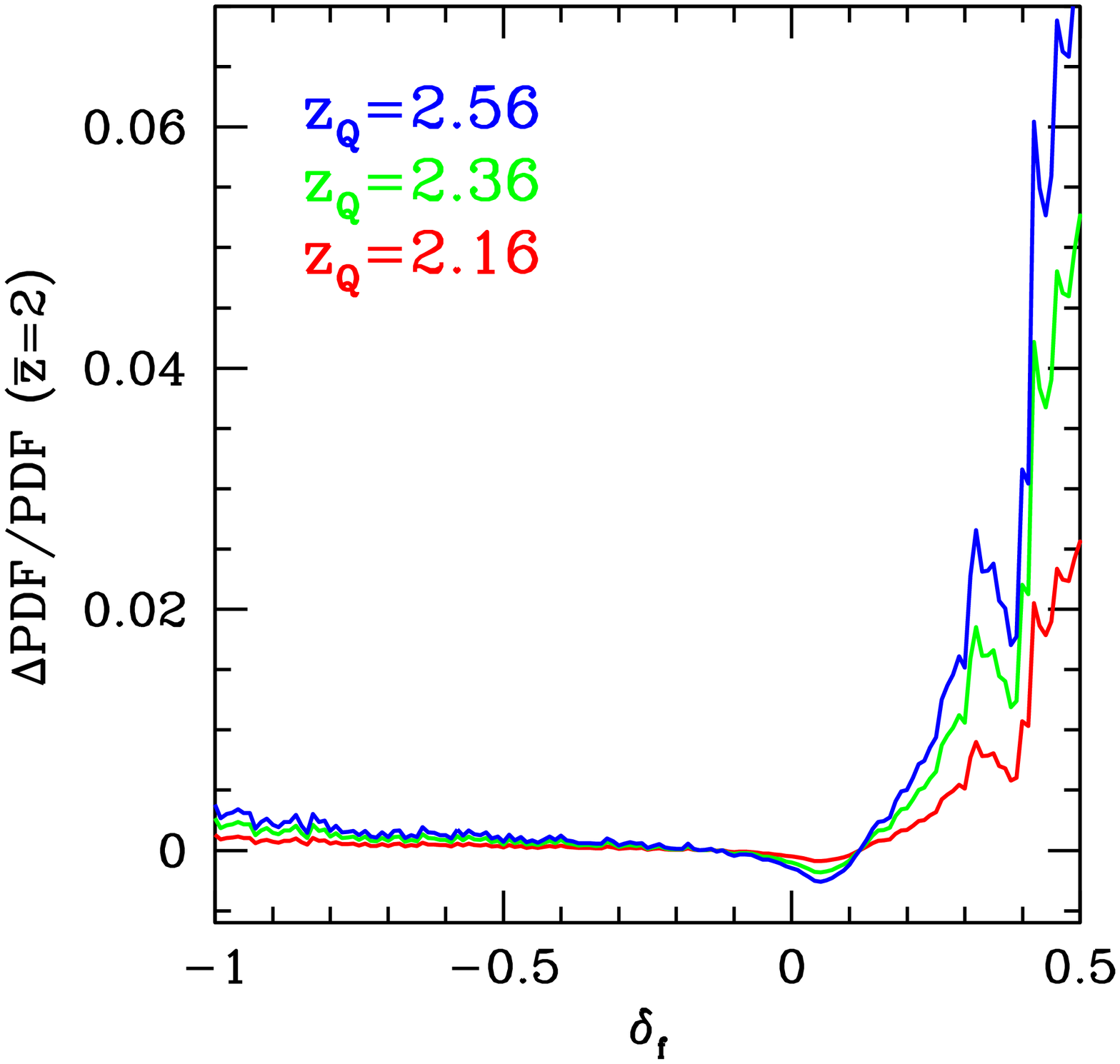}\\
\mbox{(a)}&\mbox{(b)}\\
\includegraphics[width=0.45\textwidth]{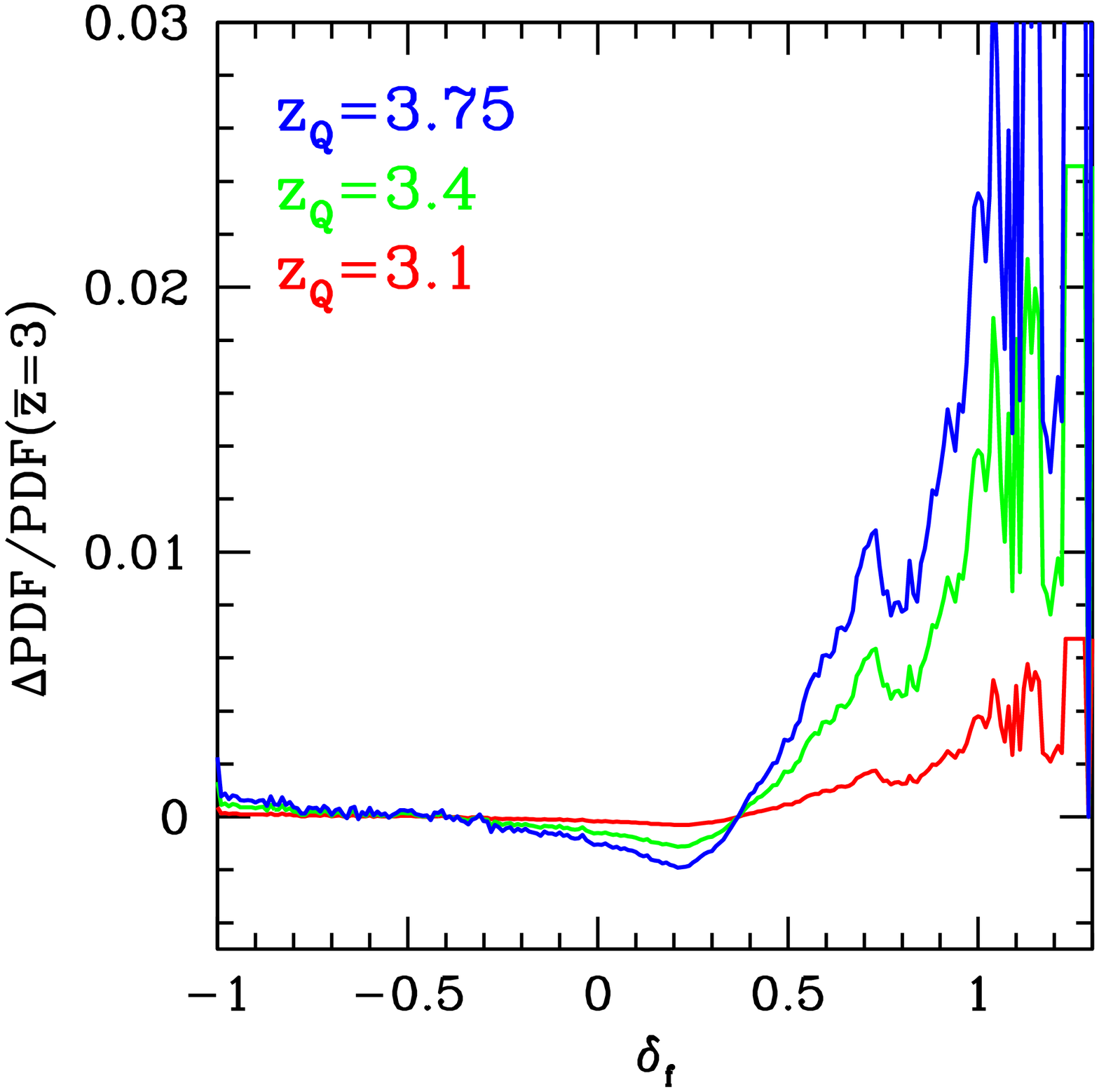}&\includegraphics[width=0.45\textwidth]{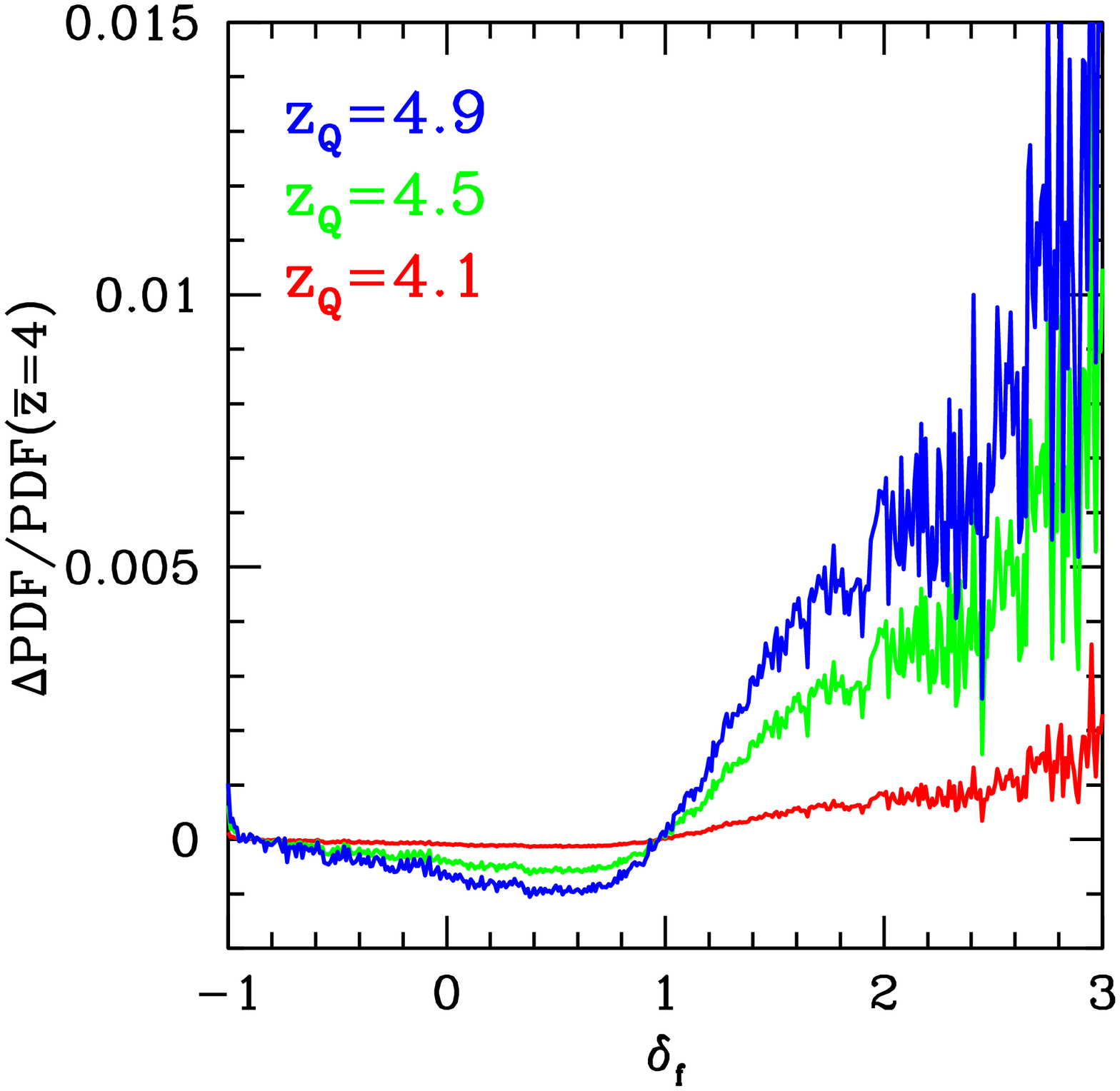}\\
\mbox{(c)}&\mbox{(d)}
\end{array}$
\caption{\label{Fig:PDFdeltaSims}(a) The probability distribution function for $\delta_f$ as measured from the simulations. Panels $(b)$ - $(d)$ show the fractional correction from magnification for mean redshifts $\bar{z}=2, 3, 4$. 
}
\end{center}
\end{figure}

Another interesting statistic of the Lyman-$\alpha$ forest is the flux probability  distribution function (PDF): $\mathcal{P}(f) df$ describes the probability  that the observed flux lies between $f-df/2$ and $f + df/2$. Here $f$ is the continuum-normalized flux (i.e. $f/f_c$ from the previous section). 

The flux PDF measured from simulations with bins of $d f=0.01$ is shown in panel $(a)$ of Fig. \ref{Fig:PDFetauSims}.\footnote{We use the same fake continuum fitting procedure described in footnote \ref{continuumfootnote}, but again this appears to make little difference to the size of the fractional lensing correction.} The fractional correction from lensing magnification is shown in panels $(b)$--$(d)$ of the same figure. The effect of magnification is to boost the low flux (high absorption) end of the PDF and decrease the PDF at the high flux end. The correction is $< 1\%$, so unimportant at the current level of precision ($5-10\%$ on the flux PDF in bins of $d f=0.05$ \cite{Kim:2007rr} but see also  \cite{McDonald:1999dt,Desjacques:2006pe,Becker:2006qj}).

It has been suggested that to avoid systematic errors from continuum fitting one should measure the PDF for fluctuations in the flux $\delta_f=f/\bar{f}-1$ rather than measuring the PDF of $f$  \cite{Lidz:2006zz}. In this case the flux is normalized along each line of sight, so the range of $\delta_f$ for one line-of-sight is $-1$ to $1/\bar{f}-1$ where $\bar{f}$ is the mean flux along that line of sight. The range of the entire PDF determined from many lines of sight will be $-1$ to $1/\bar{f} {}_{\rm min}-1$, where $\bar{f} {}_{\rm min}$ is the mean flux along the line of sight with the lowest transmission. 

The PDF for $\delta_f$ is shown in panel $(a)$ of Fig. \ref{Fig:PDFdeltaSims}.  Again the fractional correction due to lensing is shown in panels $(b)$--$(d)$. In this case the correction due to lensing appears rather large at high $\delta_f$. The high $\delta_f$ bins are dominated by lines of sight with the lowest mean flux, which are also most highly magnified. However, these high $\delta_f$ pixels are quite rare, i.e. the PDF at high $\delta_f$ is quite small and therefore noisy from an observational point of view.


\label{FluxPDF}

For an analytical estimate we apply the fundamental Eq. (\ref{Eq:MagBiasFinal}), 
\ba
\label{Pfbias0}
\langle \mathcal{P}(f) {}_{\rm obs} \rangle - \mathcal{P}(f) {}_{\rm true} = \langle \mathcal{P} (f) \delta^\mu \rangle
\ea
where $\mathcal{P} (f) df$ is estimated from each line of sight in the standard fashion: counting the fraction of pixels with a flux that falls within $df$ of $f$. The flux is a non-linear function of the gas density, but for simplicity, we will assume that gas traces mass and that $f$ is a local function of the mass fluctuation $\delta$, i.e. $f = F(\delta)$, where the function $F$ is to be specified. Let us use the symbol $p$ to denote the  (average) location of interest, i.e. $\delta_p$ is the mass fluctuation at point $p$. Thus, the flux PDF at $p$ can be expressed in terms of the mass PDF $\mathcal{P}_m (\delta_p)$:
\ba
\mathcal{P} (f) = \int d\delta_p \, \mathcal{P}_m (\delta_p) \, \delta_D (f - F(\delta_p)) = \int d\delta_p d\delta_l \, \mathcal{P}_m (\delta_p, \delta_l) \, \delta_D (f - F(\delta_p))
\ea
where $\delta_D$ is the Dirac delta function. For the second equality, we have introduced $\delta_l$ which is the mass fluctuation at some other point (the subscript $l$ is
used in anticipation of the fact that this will be where some lens is), and $\mathcal{P}_m(\delta_p, \delta_l)$ is the joint mass PDF at the two points, i.e. the one-point PDF
$\mathcal{P}_m (\delta_p)$ is related to the two-point PDF by $\mathcal{P}_m (\delta_p) = \int d\delta_l \mathcal{P}_m (\delta_p, \delta_l)$. The motivation for introducing the two-point mass PDF is to ease the computation of $\langle \mathcal{P}(f) \delta^\mu \rangle$, where $\delta^\mu$ involves a line-of-sight integral over locations that generally differ from $p$. We obtain: \footnote{Here, we are abusing the notation a bit. On the left, $\mathcal{P}(f)$ strictly speaking denotes a stochastic quantity: it should be the estimator that simply counts the fraction of relevant pixels, whereas on the right, $\mathcal{P}_m (\delta, \delta')$ denotes the true (non-stochastic) two-point mass PDF.}
\ba
\label{Pfbias}
\langle \mathcal{P}(f) \delta^\mu \rangle = \frac{3}{2}H_0^2\Omega_m (5s-2) \int_0^{\chi_Q} d\chi_l \frac{\chi_Q-\chi_l}{\chi_Q}\chi_l (1+z_l) \int d\delta_p d\delta_l\, \delta_l
\mathcal{P}_m (\delta_p, \delta_l) \delta_D (f - F(\delta_p)) \, .
\ea
We need a model for the two-point mass PDF. In linear theory, that is completely specified by the two-point correlation  $\langle \delta_p \delta_l \rangle$. The relevant fluctuations in the forest are not quite linear. Instead, we will adopt the lognormal model (see for example \cite{Kofman:1993mx,Coles:1991}):
\be
\label{Pmlognormal}
\mathcal{P}_m (\delta_p, \delta_l) =\frac{1}{2\pi\sqrt{\textrm{det\,C}}\,(1+\delta_p)(1+\delta_l)}e^{-\frac{1}{2} {\bf \Delta}^T {\bf C}^{-1} {\bf \Delta}}
\ee
where ${\bf \Delta}$ is a 2-component vector with $\Delta_i = \ln(1+\delta_i)+\frac{1}{2}\ln(1+\langle\delta_i^2\rangle)$, and ${\bf C}$ is a $2 \times 2$ matrix with components ${\rm C}_{ij}=\ln(1+\langle\delta_i\delta_j\rangle)$. Here $i$ and $j$ stand for $p$ or $l$. The lognormal model is simply one in which $\Delta$ is a Gaussian random field, and the observed $\delta$ is related to it by $1 + \delta = {\,\rm exp\,}[\Delta - \langle \Delta^2 \rangle/2]$, such that $\langle \Delta^2 \rangle = {\,\rm ln\,} (1 + \langle \delta^2 \rangle)$.
Substituting Eq. (\ref{Pmlognormal}) into Eq. (\ref{Pfbias}) and integrating over $\delta_p$ and $\delta_l$, we find
\ba
\label{Eq:PDFLogNorm}
{\langle \mathcal{P}(f)_{\rm obs} \rangle - \mathcal{P}(f)_{\rm true} \over \mathcal{P}(f)_{\rm true}} = \frac{3}{2}\Omega_m(5s-2) H_0^2 \int_0^{\chi_Q} d\chi_{l}
{\chi_Q - \chi_l \over \chi_Q} \chi_l (1+z_l) \left(e^{C_{lp}/C_{pp}\ln(1+\delta_*)+\frac{1}{2}(C_{lp}-C_{lp}^2/C_{pp}))}-1\right)\, ,
\ea
where $\delta_* \equiv F^{-1} (f)$ is the actual density such that the observed flux equals the value of interest $f$. At this point, we need to specify $F$: We use $F(\delta) = {\,\rm exp\,}[ - A (1 + \delta)^\beta]$, with $\beta = 1.58$ and $A=0.2010, \, 0.9578,\, 2.960$ at $z_p=2,\,3,\,4$ respectively. This approximates well what is in our simulations, though the simulated spectra were computed using the exact relation between the optical depth, baryon density, temperature and ionizing background.

\begin{figure}
\begin{center}$\begin{array}{cc}
\includegraphics[width=0.45\textwidth]{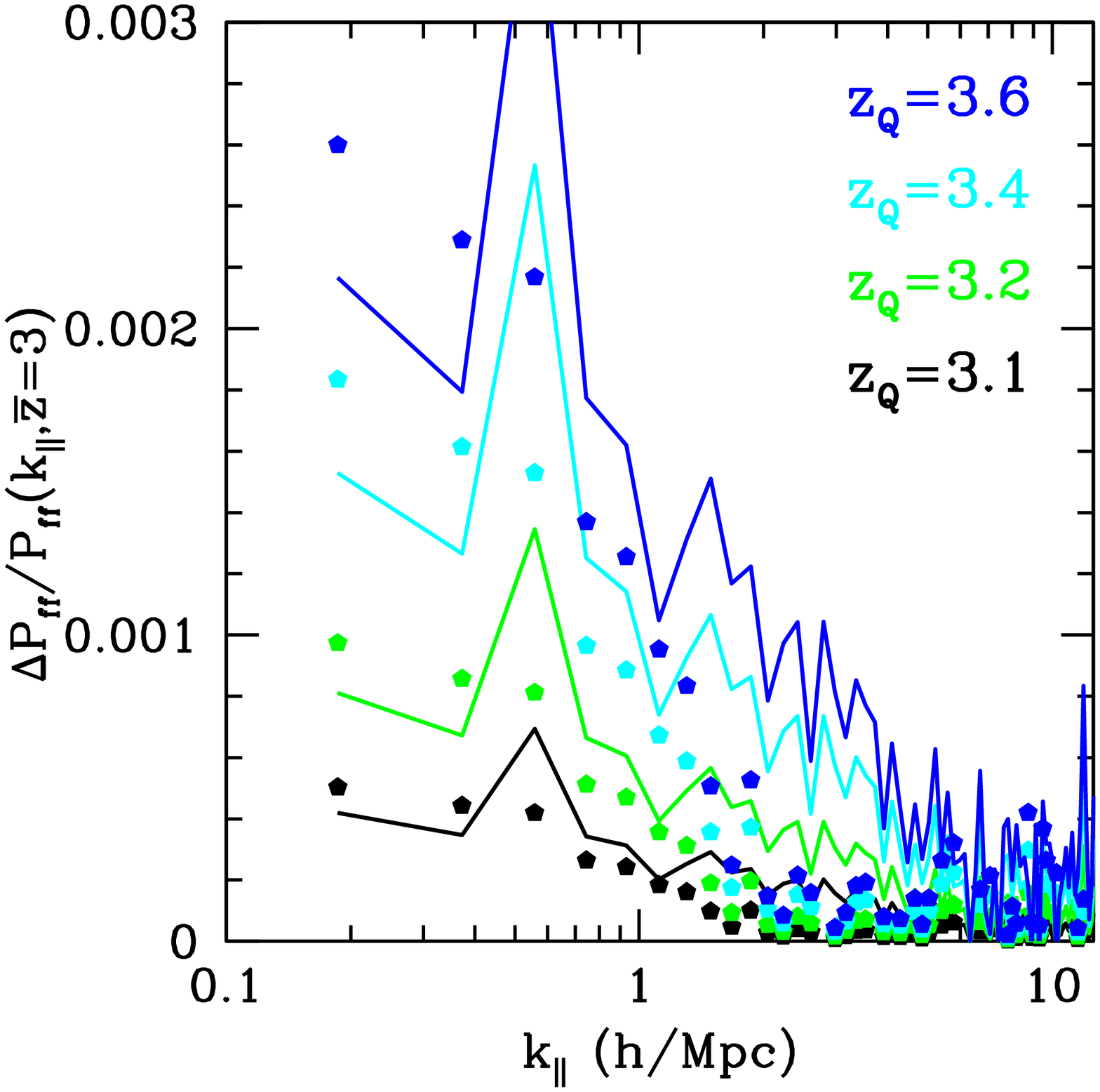}&\includegraphics[width=0.45\textwidth]{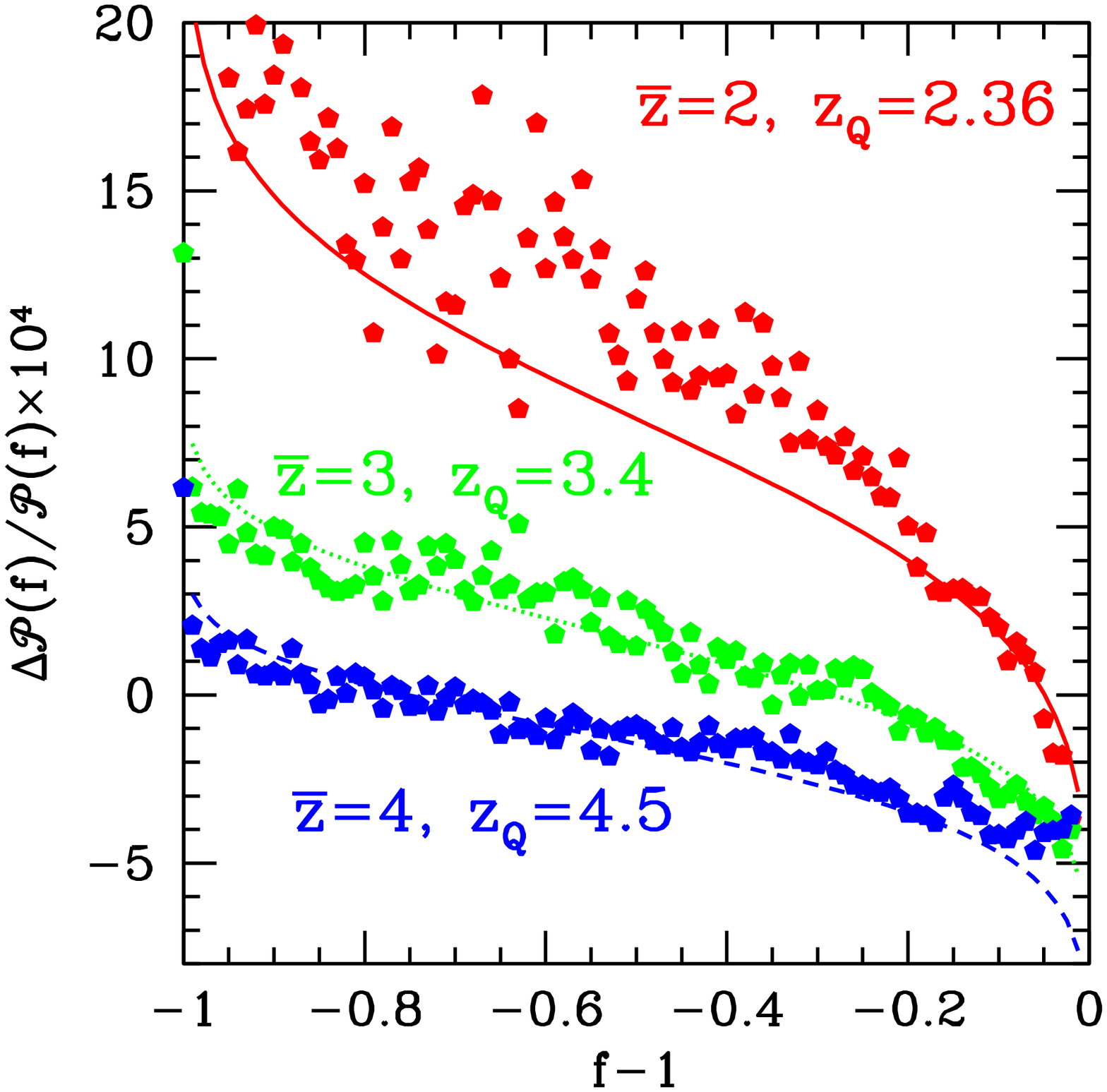}
\end{array}$
\caption{\label{Fig:PffBispec} \label{Fig:PDFLogNorm} Left: A comparison of the predicted lensing bias to the flux power spectrum using the weighted calculation presented in \S \protect\ref{Sims} (solid lines) versus using Eq. (\protect\ref{Eq:DPBffmHighK}) with the flux-flux-mass bispectrum measured from simulations (points). From bottom to top the quasar redshifts are $3.1, 3.2, 3.4, 3.6$.  Right panel: The correction to the flux PDF from the analytic model described in Eq. (\ref{Eq:PDFLogNorm}) (solid lines) compared with the results from simulations (points). The analytic description with the log-normal model for the mass PDF works surprisingly well.}
\end{center}
\end{figure}

To use Eq. (\ref{Eq:PDFLogNorm}), we need $C_{lp}$ and $C_{pp}$, which we compute using the nonlinear mass power spectrum, with suitable smoothing (for $p$) to account for the effective Jeans smoothing of baryons. The cosmology and power spectrum prescriptions are described at the end of \S \ref{LensBiasDeriv}. Figure \ref{Fig:PDFLogNorm} compares the analytic calculation in Eq. (\ref{Eq:PDFLogNorm}) with the results from simulations in \S \ref{Sims}. The two methods agree remarkably well considering the simplicity of the analytic method. If the linear mass PDF were used, the calculated correction has a slightly different $f$-dependence and is typically smaller by about a factor of $5-6$. 

We should mention data that do not resolve the Jeans scale (or roughly, the thermal broadening scale) have an additional complication: even if the fundamental mass density
is lognormal distributed (in both one-point and two-point sense),  the flux field smoothed with a coarse resolution might not be well described by our model. In other words,
smoothing and nonlinear transformation do not commute. In any case, the lensing effect on the flux PDF appears to be fairly small.

\section{Lensing as Signal: a test of how neutral hydrogen traces mass}
\label{MagCorrelation}

In the previous section, we have been exploring lensing as a source of measurement bias. Here, we explore the converse: lensing as a useful signal. We are interested in ways to extract signals of gravitational lensing from Lyman-alpha forest observations and thereby constrain the cross-correlations between flux and mass (for example, the flux-mass power spectrum and flux-flux-mass bispectrum present in Eq. (\ref{Eq:DPBffmHighK}) and Eq. (\ref{Eq:TauBias})). 

One option is to correlate the magnitude of quasars with Lyman-alpha forest observables (we will consider other possible cross correlations at the end of this section). Given some observable $\justO_I$ measured from a quasar labeled $I$,  and its magnitude $m_I$, we can form an estimator $\mathcal{E}$:
\ba
\label{EstimatorE}
\mathcal{E} = {1 \over N_{\rm QSO}} \sum_I m_I \justO_I -  {1\over N_{\rm QSO}^2} \left(\sum_I m_I\right) \left(\sum_I \justO_I\right) 
\ea
where $N_{\rm QSO}$ is the number of quasars in one's sample. As is explained in \S \ref{LensBiasDeriv}, the Lyman-alpha forest observable is always implicitly weighted by the number density of quasars (determined by the sample selection). And the quasar magnitude is of course modified by lensing, following from Eq.  (\ref{Eq:fmag}):
\be 
\label{deltam} \delta m = -5 \kappa / {\rm ln\,} 10\, .  
\ee 
It is shown in Appendix \ref{derive} that
\ba
\label{Eaverage}
\langle \mathcal{E} \rangle = 5 \tilde s \langle \kappa \delta\mathcal{O} \rangle
\ea
where $\delta\mathcal{O}$ represents fluctuations in $\mathcal{O}$  and $\tilde s$ is defined as
\ba
\label{stilde}
\tilde s \equiv {1\over {\rm ln\,} 10} {\int dm \, \epsilon(m) (dn^0 /dm) (m - \bar m) \over \int dm \, \epsilon(m) n^0(m)}
\ea
where $\bar m$ is the average magnitude in the sample i.e. $\bar m = \int dm \, m \epsilon(m) n^0(m)/ \int dm \, \epsilon(m) n^0(m)$, and $n^0(m)$ and $\epsilon(m)$ are the luminosity function and selection function respectively --- the definition for $\tilde s$ is chosen to resemble those for $s$ and $s'$ (Eqs. [\ref{nlensed}], [\ref{sprime}]). The $\tilde s$ defined here is related to $C_S$ defined in \cite{Menard:2009yb} by $C_S = - {\rm ln\,} 10 \, \tilde s$. As first emphasized by \cite{Menard:2009yb}, $C_S = 0$ for a strictly power-law luminosity function in the sense of $n^0 (m) \propto e^m$. The realistic quasar luminosity function is not of this form, and we will adopt $C_S \sim 1/3$ from \cite{Menard:2009yb}, or equivalently $\tilde s \sim - 0.14$, for our numerical estimates below. The normalizing factor in Eq. (\ref{Eaverage}) is therefore $5 \tilde s \sim - 0.7$. It is worth emphasizing that there is no `-2' term here, unlike for instance $5s-2$ in Eq. (\ref{Eq:MagBiasFinal}). The `-2' there arises from the geometrical increase in area by magnification. The statistic $\mathcal{E}$ under consideration is immune to this effect. Further discussions can be found in Appendix \ref{derive}.

\subsection{Magnitude-flux Correlation}
\label{magflux}

Let us first consider cross-correlations between ${\cal O} = \delta_f$, the intervening flux fluctuation, and the quasar magnitude $m$, giving
\ba
\label{magfluxexpect}
\langle \mathcal{E}_{\delta m \, \delta_f} (\chi_Q, \chi) \rangle = 5 \tilde s \langle \kappa \delta_f \rangle \approx {3\over 2} H_0^2 \Omega_m 5 \tilde s {\chi_Q - \chi \over \chi_Q} \chi (1+z) P_{fm}(k_\parallel=0)
\ea
where $\chi_Q$ is the distance to quasar, $\chi$ is the distance to absoprtion (i.e. where $\delta_f$ is located) and $z$ is the corresponding redshift. Here, $P_{fm}$ is the (true) 1D flux-mass power spectrum  --- the subscript $m$ of $P_{fm}$ stands for mass rather than magnitude (just as in \S \ref{TauCorrection}).

The nice thing about the estimator $\mathcal{E}_{\delta m \, \delta_f}$ is that it is expected to depend on the location of absorption $\chi$ in a predictable way, which allows this to be separated from other possible systematic effects, an example of which is large scale power from the uncertain continuum (shape).  The fact that the amplitude of this  cross correlation signal scales with $\tilde s$ can also be exploited to test for consistency, for instance by isolating different subsamples of quasars with different values of $\tilde s$.

In Fig. \ref{Fig:deltafXmagSDSS} we show the expected signal for this  cross-correlation as a function of distance to the quasar $ \chi_Q - \chi$, where $\chi$ is the distance to the absorption. To estimate the overall statistical significance, we can combine the measurements at different separations into an estimate for a single amplitude, namely $P_{fm}(k_\parallel=0)$. The appropriate minimum variance estimator is described in Appendix \ref{estimator}, where we also derive its signal-to-noise:
\ba
\label{magfluxS2N}
{S \over N} = \sqrt{ {N_{\rm QSO} \over \langle \delta m^2 \rangle} \int {dk_\parallel \over 2\pi}  {|\mathcal{E}_{\delta m \delta_f} (k_\parallel)|^2  \over P_{ff}(k_\parallel) +\mathcal{S.N.}}}
\ea
where $\mathcal{E}_{\delta m \delta_f} (k_\parallel)$ is the Fourier transform (over $\chi$) of Eq. (\ref{magfluxexpect}), $N_{\rm QSO}$ is the number of quasars available, $\langle \delta m^2 \rangle$ is the quasar magnitude variance, $P_{ff}$ is the 1D flux power spectrum, and $\mathcal{S.N.}$ is the associated shot-noise. 

The $\mathcal{S.N.}$ term is important so we consider the signal-to-noise per quasar for two survey configurations: `SDSS III configuration' with resolution FWHM is $60$ km/s, and the shot-noise power is $\sim 0.44$ Mpc/h $(3/(1 + \bar z))^{3/2}$ \cite{Hui:2000rw} and `Keck configuration' with resolution FWHM is $10$ km/s and shot-noise power is $\sim 0.029$ Mpc/h $(3/(1 + \bar z))^{3/2}$. For both we take the magnitude dispersion of the quasar sample to be $0.5$ \cite{Menard:2009yb}. The first set of numbers roughly resemble the expectations of SDSS III \cite{SDSSIII} but keep in mind the configuration we assume likely differs a bit from what will turn out in practice. The results for the $S/N$ per sight-line are summarized in Table I. For the SDSS III expectation of $N_Q=160,000$ the ${S/ N}$ for this amplitude is  expected to be only $\sim 1$ and even for a futuristic survey like BigBOSS \cite{Schlegel:2009uw} that could measure $10^6$ spectra with similar resolution only ${S/N} \sim 2-3$ could be achieved. Comparison with the `Keck configuration' shows that shot noise is clearly a limitation for these observables. However, the $S/N$ estimates presented are for a single redshift bin of width $\Delta z\sim0.2$, but measurements will be made at a range of redshifts which could be combined to get a slightly higher signal-to-noise estimate of, for example, the mean $P_{fm}$ across the redshift span of the survey.  

It is worth noting a few points that reduce $S/N$ of the forest-magnitude correlations in comparison with galaxy-magnitude correlations used \cite{Menard:2009yb}. (1) The cross-correlation between flux and mass is lower than the correlation between galaxies and mass (the correlation coefficient is $\sim P_{fm}/\sqrt{P_{ff}P_{mm}}\lsim 0.5$).  (2) Lensing peaks at a distance halfway between the observer and the quasar, but to avoid confusion with Lyman-$\beta$ we use only the part of the Lyman-$\alpha$ forest near to the quasar where the lensing is smaller. (3) The signal-to-noise of magnitude-forest correlations is $\propto \sqrt{N_{Q}}$ where $N_Q$ is the number of quasars with spectra, while for the galaxy-magnitude cross-correlation it is proportional to the number of galaxy-quasar pairs $\propto \sqrt{N_Q N_g}$  
(however in \S\ref{othercorrelations} we briefly discuss correlating the forest with quasars along different lines of sight).

\subsection{Magnitude-Power Correlation}
\label{magpower}

\begin{figure}
\begin{center}
$\begin{array}{cc}
\includegraphics[width=0.45\textwidth]{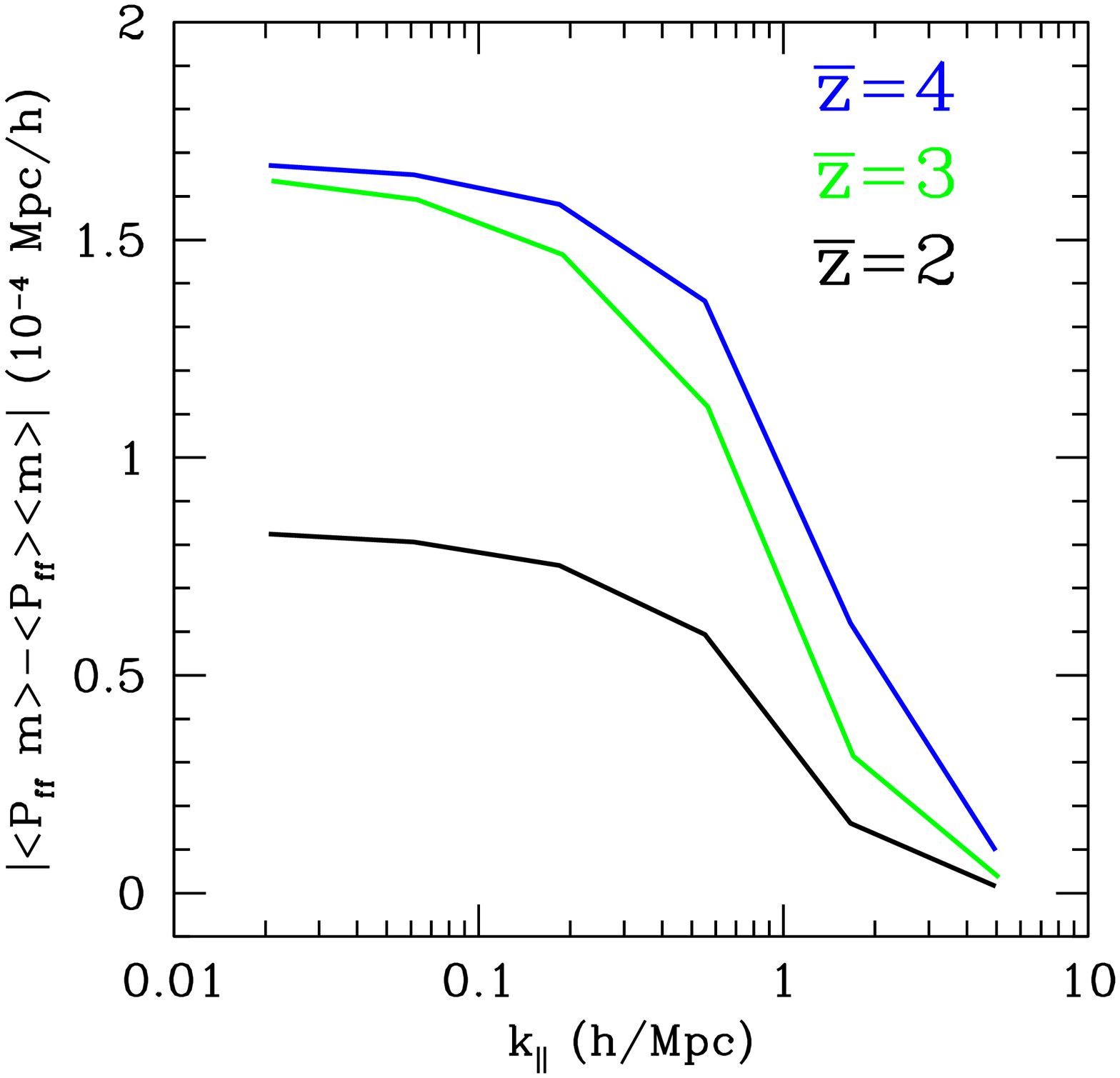} & \includegraphics[width=0.45\textwidth]{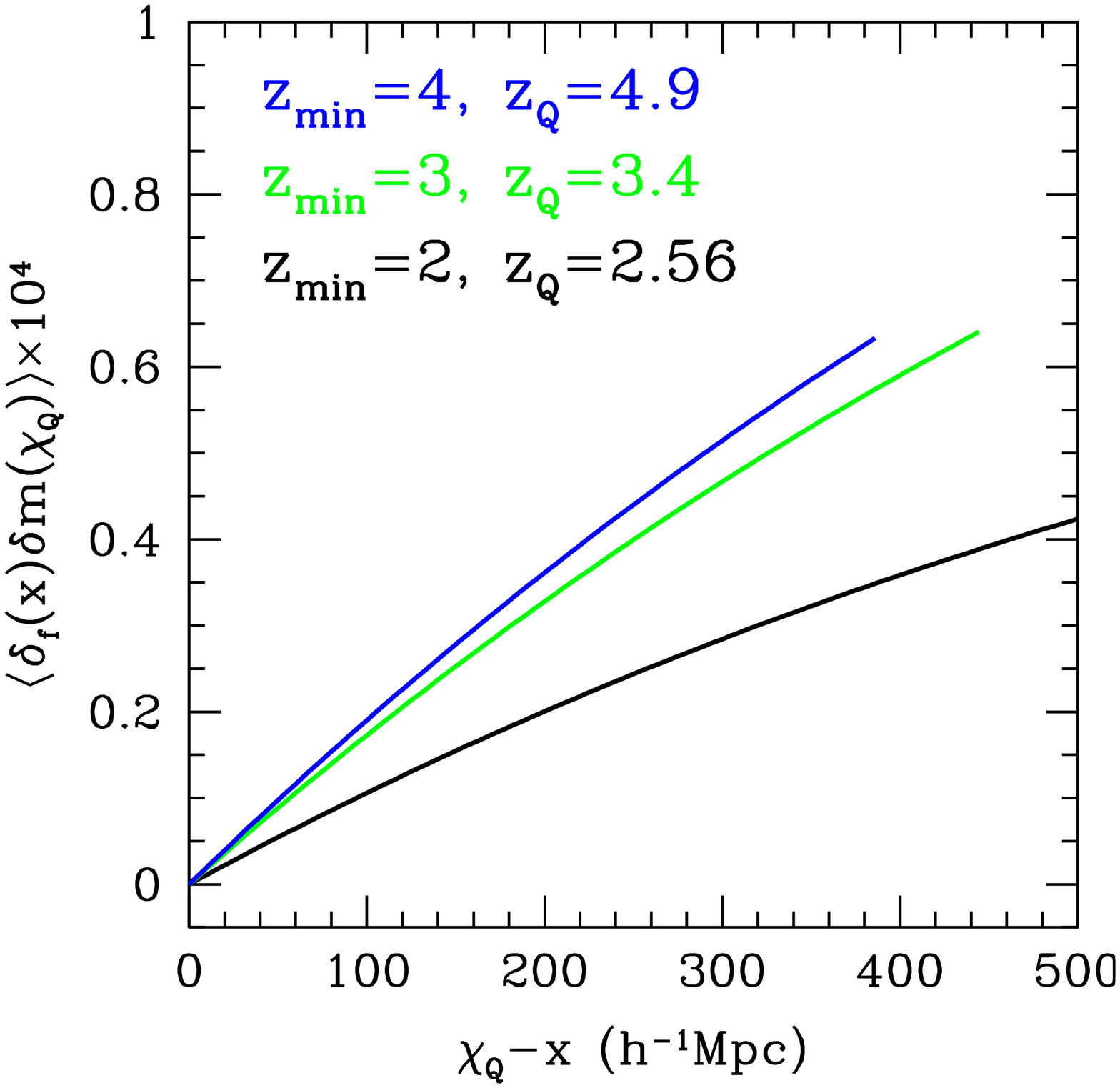}
\end{array}$
\caption{\label{Fig:PffMag}\label{Fig:deltafXmagSDSS}Left: The lensing-induced correlation between quasar magnitude and the flux power spectrum given in Eq. (\ref{magpowerexpect}). The amplitude of the signal increases with increasing distance between $\chi(\bar{z})$ and the background quasar but above we assume the correlation is averaged over background quasars at redshifts between $\bar{z}+0.1$ and $z_\beta=\nu_\beta/\nu_\alpha(1+\bar{z})-1$ (the maximum redshift for the quasar before confusion between Lyman-$\alpha$ and Lyman-$\beta$ absorption sets in).  Right: The magnitude-flux cross correlation (Eq. \ref{magfluxexpect}) as a function of line-of-sight comoving distance to the quasar. }
\end{center}
\end{figure}
\begin{table}\label{StoNtable}
\begin{tabular}{|c|c|c|}
\hline
Observable: & $S/N$ per l.o.s. for SDSS III configuration & $S/N$ per l.o.s. for Keck configuration \\
\hline 
$\langle \delta_f\delta m\rangle$ at $z=2,3,4$ & $0.002, 0.0029, 0.0022$ &  $0.0041, 0.0036, 0.0024$ \\
\hline
$\langle \delta P_{ff}\delta m\rangle $ at $z=2,3,4$ & $0.0016, 0.0028, 0.0025$ & $0.008, 0.006, 0.0042$ \\
\hline 
$\langle f/f_C\delta m\rangle$ at $z=2,3,4$ & $0.0054,0.0032,0.00095$ & --- \\
\hline
\end{tabular}
\caption{The signal-to-noise per sightline in redshift bins of $\Delta z \sim 0.2$ for the proposed estimators  in Eq. (\ref{magfluxexpect}), (\ref{magpowerexpect2}) and (\ref{eq:fmcorr})  for the SDSS III and Keck survey configurations (see \S\ref{magflux}). Here we assume the quasar sample has $\tilde{s}=-0.14$ (see Eq. (\ref{stilde})) and the quasar magnitude dispersion is $0.5$. For $\langle f/f_C\delta m\rangle$ we assume the same fractional errors of $f/f_C$ as \cite{Bernardi:2002mr} and that the errors are statistics limited so they scale as $1/\sqrt{N_{QSO}}$.  }
\end{table}

Another possibility is to set $\mathcal{O} = P_{ff}(k_\parallel)$  in the estimator Eq. (\ref{EstimatorE}), i.e. cross correlate quasar magnitude and the flux power spectrum. This estimator has the following expectation value: 
\begin{eqnarray}
\label{magpowerexpect}
\langle \mathcal{E}_{\delta m\, P_{ff}} (k_\parallel) \rangle
= {5 \tilde s \over \Delta\chi}
\langle \kappa \delta_f (k_\parallel) \delta_f^* (k_\parallel)
\rangle
\end{eqnarray}
where $\Delta\chi$ is the length of the quasar spectra from which the power is measured. This expression is similar to Eq. (\ref{Pffbias}), and indeed one can relate this to the 
bias in $P_{ff}$ measurement we have calculated in \S \ref{PCorrection} (Eq. [\ref{Eq:DPBffmHighK}]):
\begin{eqnarray}
\label{magpowerexpect2}
\langle \mathcal{E}_{\delta m\, P_{ff}} (k_\parallel) \rangle
= {5 \tilde s \over 5s - 2}
(\langle P_{ff} {}_{\rm obs} (k_\parallel) \rangle
- P_{ff} {}_{\rm true} (k_\parallel)) \, .
\end{eqnarray}
This means that one could in principle use the magnitude-power correlation to correct for the bias in $P_{ff}$ measurement. The expected magnitude-power correlation is shown in Fig. \ref{Fig:PffMag} and the signal-to-noise is given in Table I.

\subsection{Magnitude-Mean-Transmission Correlation}
\label{magmeanflux}
If the continuum can be accurately estimated, one could also correlate the mean transmission from each line of sight with the quasar magnitude behind it, i.e. choose $\mathcal{O} = [f/f_C]$ where $f$ is the observed flux and $f_C$ is the continuum, and the brackets $\, [\,\,]\,$ denote an average along the line of sight. One can see that the expected correlation should be related to the bias in $e^{-\tau_{\rm eff}}$ discussed in \S \ref{SimsTau}: 
\begin{eqnarray}
\label{eq:fmcorr}
\langle \mathcal{E}_{\delta m \, [f/f_C]} \rangle ={5 \tilde s \over 5 s - 2} ( \langle e^{-\tau_{\rm eff}} {}_{\rm obs} \rangle - e^{-\tau_{\rm eff}} {}_{\rm true} )
\end{eqnarray}
The signal-to-noise per sight line is listed in Table I. Note that the number of quasars needed is realistic only if the continuum is estimated by extrapolating from the red side (such as in the analysis of \cite{Bernardi:2002mr}).  The number of quasars should be much smaller if the continuum were estimated by performing a smooth fit through portions of the Lyman-alpha forest that are deemed unabsorbed, a method that typically requires high resolution data. Whichever the continuum estimation method, care should be taken in accounting for possible systematic biases (in the continuum estimate) when interpreting this cross correlation measurement.

\subsection{Additional Cross Correlations}
\label{othercorrelations}
We have focused on cross correlations between the quasar magnitude and a Lyman-alpha forest observable {\it in the same line of sight}. There are two possible extensions we should mention in passing. One is that the quasar magnitude and the Lyman-alpha forest observable can come from {\it different lines of sight}. In other words, the cross correlations can be measured at a non-zero angular separation. On the one hand, the forest-magnitude correlation should drop off rapidly with increasing angular separation decreasing the signal, however the noise is reduced by the number of quasar pairs at a given angular separation. Roughly, the signal-to-noise for lines of sight at angular separation $\theta$ should scale as
\be
\left(\frac{S}{N}\right)_{\textrm{{\small at sep.}}\theta}\sim \left(\frac{S}{N}\right)_{\textrm{{\small same l.o.s.}}}\frac{10(\chi_Q-\bar{\chi})}{\chi_Q}\frac{w_{fm}(\theta)}{w_{fm}(\theta=0)}\sqrt{\frac{N_{pairs}(\theta)}{N_Q}}
\ee
where $w_{fm}(\theta)$ is the angular flux-mass correlation function, $N_{pairs}(\theta)$ is the number of quasar pairs with angular separation $\theta$ and the term with the ratio of the distances roughly accounts for the fact that this correlation doesn't require $\bar{z}$ to be above the Lyman-$\beta$ confusion limit for the quasar (previously we needed $\bar{\chi}>\chi_\beta \sim 4/5\chi_Q$ and had set $\bar{\chi}\sim \chi_\beta+ (\chi_Q-\chi_\beta)/2$). It seems that the sparseness of quasars ($\sim 16/(\textrm{deg.})^2$ for SDSS III) does not permit $N_{pairs}(\theta)$ to be large enough that adding off-axis correlations will drastically improve the signal-to-noise in the near term.

Another possibility is to cross correlate the quasar number density (which is affected by lensing through magnification bias) with the Lyman-alpha forest observable. Such a correlation makes sense only at a non-zero lag (or non-zero smoothing),  since the Lyman-alpha forest is observable only  if there is a quasar directly behind it.  While we do not give explicit estimates here of the signal-to-noise for these cross correlations, they should be useful in disentangling the lensing signal from certain systematic effects, as we will discuss next.

\subsection{Dust and Other Systematic Effects}
\label{dust}

We discuss here three systematic effects that could complicate the
measurement of the various cross correlations mentioned above.

The first is the continuum. The continuum presumably is smooth and therefore
has fluctuations only on large scales. But since its precise shape is uncertain,
a cross correlation such as the magnitude-flux correlation is susceptible
to possible contamination from continuum power.
A fortunate feature of the cross correlation is that it has a definite
shape predicted by lensing, as well as an amplitude that scales with $\tilde s$.
Both can be exploited to check for such a contamination.

The second systematic effect we loosely refer to as `background subtraction'.
Realistic spectra of quasars inevitably contain `background' which can come
from several sources, including the sky and scattered light within the 
optical instrument. While attempts are generally made to subtract these backgrounds
as accurately as possible, there are inevitably residuals. These residuals
could correlate with the quasar magnitude, for instance they could be more
noticeable for fainter quasars. This would then produce spurious correlations
when we correlate the quasar magnitude with Lyman-alpha forest observables (deduced
from imperfect data that contain residual backgrounds). 
In fact, existing flux power spectrum measurements from SDSS \cite{McDonald:2004eu} 
are known to exhibit an otherwise puzzling correlation: that $P_{ff}$ is systematically higher
for fainter quasars, and this correlation is statistically significant
(the normalized correlation coefficient is $\sim 5 - 10\%$ \cite{Abazajian:2010}). 
Such a correlation can be explained by this background subtraction effect.
It cannot be explained by lensing since it tends to produce a correlation of an opposite
sign unless $\tilde s$ has a sign opposite to what is known.
(It can also be plausibly produced by dust which we will discuss below).
To disentangle background subtraction issues from lensing, it would be useful
to examine magnitude-observable cross correlations at non-zero lag -- it would
be quite surprising if background residuals cause correlations between quasar
magnitude at one point with Lyman-alpha forest power at another.

The third systematic effect is dust extinction. Dust modifies flux by
$f \rightarrow f e^{-\tau_{\rm dust}}$, where $\tau_{\rm dust}$ is the optical depth due
to dust. This modification is frequency dependent (higher optical depth for bluer photons),
but the frequency (which translates into scale) dependence is mild on the scales of interest, and therefore effectively acts as a continuum.
The net magnification plus dust correction to the quasar number density is
\be
{n}\rightarrow n\, {\mu}^{2.5s-1} e^{-2.5 s {\tau}_{\rm dust}} \, .
\ee
Including dust extinction, the measurement bias associated with a
Lyman-alpha forest observable (Eq.  [\ref{Eq:MagBiasFinal}]) is changed to 
\be 
\label{Eq:MagBiasFinaldust}
\langle\mathcal{O}_{\rm obs} \rangle
- \mathcal{O}_{\rm true} 
= (5s-2) \langle \mathcal{O} \kappa \rangle - 2.5 s \langle \mathcal{O} \delta\tau_{\rm dust} \rangle
\ee
where ${\delta\tau}_{\rm dust}=\tau_{\rm dust}-\langle \tau_{\rm dust}\rangle$, which is assumed
to be small, and we have ignored $\langle \mathcal{O} \kappa \delta\tau_{\rm dust} \rangle$.

As far as the `signal' part of our discussion is concerned,
the magnitude-observable cross correlation (Eq. [\ref{Eaverage}])
is modified to
\begin{eqnarray}
\label{Eaveragedust}
\langle \mathcal{E} \rangle = 5 \tilde s
\langle \kappa \delta\mathcal{O} \rangle - 2.5 \tilde s
\langle \delta\tau_{\rm dust} \delta\mathcal{O} \rangle\, .
\end{eqnarray}

A full calculation of the effect of dust is beyond the scope of this paper. 
But given a model for $\delta\tau_{\rm dust}$ the effect can be calculated in a way very similar to 
lensing -- $\tau_{\rm dust}$ after all is another line of sight integral
over the (dust) density.
Calculations comparing the amplitudes of magnification and dust corrections to supernova flux have shown $\langle\kappa{\delta\tau}_{\rm dust}\rangle-\langle{\delta\tau}_{\rm dust} {\delta\tau}_{\rm dust}\rangle$ to be $\sim 7-40\%$ of $\langle\kappa \kappa\rangle$ at $z_{source}=1.5$ depending on the dust model \cite{Zhang:2006qm}.  At higher redshifts the dust term is expected to be less important, while the lensing effect should grow. On the other hand, recent measurements suggest that at $z=0.3$ dust extinction can cause shifts in source magnitude comparable to those caused by lensing magnification \cite{Menard:2009yb}. For the Lyman-$\alpha$ forest, which is typically measured at redshifts $z\gsim 2$,  the dust corrections are likely to be smaller than the lensing corrections but should be included in a full analysis. 

\section{Discussion}
\label{Conclusions}
We have discussed the sampling bias introduced by magnification and dust on measurements of the Lyman-$\alpha$ forest. In calculating the effect of this bias (summarized in Eq.  (\ref{Eq:MagBiasFinal})) we made the assumption that all quasar spectra are weighted equally. If, on the other hand, forest measurements were weighted by quasar flux, the effect of lensing bias would be larger than what we have found (see for example Eq.  (\ref{Eq:MagBiasFluxWeight})).  

If the quasars are weighted uniformly, we find that magnification bias leads to corrections $\lsim 1\%$ to the flux power spectrum (Fig. \ref{Fig:PffSims}), $\lsim 0.1\%$ to the effective optical depth (Fig. \ref{Fig:TauBias}), $\lsim 0.1\%$ to the flux probability distribution function $\mathcal{P}(f)$, and as large as a few $\%$ to the  probability distribution function for the flux fluctuation $\mathcal{P}(\delta_f)$ (see Fig. \ref{Fig:PDFetauSims}  and Fig. \ref{Fig:PDFdeltaSims}). These estimates assume quasar number count slope $s=1$, probably reasonable for the SDSS quasars used for Lyman-$\alpha$ forest measurements. The lensing correction varies strongly with redshift with larger corrections present at lower redshift, largely due to non-linear growth of mass fluctuations. The biases induced by magnification to the effective optical depth are significantly smaller than current error bars. For the flux power spectrum, the lensing bias is just within the current error bars, and since it leads to a systematic offset in each data point, may effect measurements of the overall amplitude of the power spectrum. At low redshift the lensing effect on the PDF of $\delta_f$ can be rather large, reaching several percent at the high $\delta_f$ end of the PDF at $z=2$, however this occurs only in regions where the PDF itself is very small $\lsim 0.1$. Lensing magnification is probably unimportant for current measurements of the flux PDF and the quantities derived from it. 

One may wonder whether including nonlinear magnification\footnote{\label{nonlinfootnote}To be precise, the true magnification $\mu=1/((1-\kappa)^2-\gamma_1^2-\gamma_2^2)$ where $\gamma_1$ and $\gamma_2$ are the shear. Throughout this paper we assume $\kappa \, , \gamma_1 \, ,\gamma_2$ are small quantities and so we keep only the first order terms in the expression for $\mu$. Non-linear magnification refers to the higher-order (in $\gamma$ and $\kappa$)  terms contributing to $\mu$  \cite{Menard:2002ia}.} -- which is more important at the small angular separations relevant for the Lyman-$\alpha$ forest -- could change our results.  We have checked that including all $\kappa$ terms contributing to $\mu$ in Eq. (\ref{nlensed}) and Eq. (\ref{eq:weightdef}), rather than just the linear term as in Eq. (\ref{Eq:KappaDef}), changes the magnification bias corrections by $\lsim 2\%$. The true nonlinear magnification of course depends on the components of the shear as well as the lensing convergence but it would be surprising if they drastically changed our (all-orders in $\kappa$) estimate of their importance\footnote{Actually, there is a further caveat here in that we didn't compute the true lensing convergence $\kappa$ because we have ignored contributions from mass fluctuations at the lower redshifts outside the box. For the non-linear terms, this shouldn't be a good approximation (e.g. $\langle\kappa\kappa\rangle$ is poorly estimated) nevertheless it would be surprising if this made an orders-of-magnitude difference in the effect of magnification bias. But perhaps this plus the ignored $\gamma$ terms could increase the importance of non-linear magnification to the $\sim 20\%$ reported by \cite{Menard:2002ia}.}. 


An additional caveat to our analysis is that the damped Lyman-$\alpha$ (DLA) systems and their associated damping wings are not accurately modeled by the simulations and mock spectra. While DLAs are rare, magnification bias could make them more abundant in survey samples. The damping wings of DLAs are known to add spurious power to $P_{ff}(k_{||})$ at the $10-20\%$ level on large scales \cite{McDonald:2005ps}. Magnification bias should increase the abundance of DLAs making the power spectrum more biased than our results suggest. Precisely how magnification affects the DLA bias deserves further investigation.

Perhaps most interesting is that lensing magnification induces a correlation between Lyman-$\alpha$ observables and the magnitude of the quasars used to measure them (Fig. \ref{Fig:PffMag}). Unfortunately even with the large number of quasar spectra the BOSS survey will obtain it will be a challenge to detect correlations between the flux power spectrum, the flux decrement or the mean transmission and quasar magnitude with high signal-to-noise (see Table I). Additionally, it is possible that lines of sight with high magnification also have more metal lines which could further complicate the analysis. Nevertheless these correlations provide a direct measure of how fluctuations in the quasar flux trace the underlying density field, for example they could \emph{directly} constrain the flux-mass power spectrum and should therefore be targeted. A more thorough analysis of how to detect the flux-magnitude and flux-power correlations is necessary, we leave this to future work. For a recent idea in a similar vein see \cite{Vallinotto:2009wa,Vallinotto:2009jx} who propose correlating lensing in the cosmic microwave background with fluctuations in the forest to extract flux-mass information.  

It is worth noting that, with a large quasar sample, it may be possible to exploit the dependence on $5s-2$ and/or $\chi(z_Q)$ to isolate the magnification correction and to measure the flux-magnitude correlations. The lensing correction depends linearly on $5s-2$, which ranges quite a bit depending on magnitude limit (Fig. \ref{Fig:slope}). In the weak lensing limit the correction scales as $H_0(\chi(z_Q)-\chi(z))\chi(z)/\chi(z_Q)$ which varies from $\sim \frac{4}{25}H_0\chi(z_Q)$ to $0$ as $z$ goes from $z_\beta$ and $z_Q$.   Indeed measuring these distinctive dependencies would help guard against possible instrumental systematics from being confused with the lensing signal.  Our estimates of the forest-magnitude correlations in \S \ref{magflux}--\ref{othercorrelations} assumed $z$ was halfway between $z_\beta$ and $z_Q$. Also, there exist in public data $\sim 150$ close quasar pairs \cite{Hennawi:2006xm}, that one might wish to use for this analysis, unfortunately their number is small compared to the $10^5$ expected from SDSS III and will not significantly improve the signal-to-noise.

There is quite a bit of interest in using the Lyman-$\alpha$ forest to map the 3D density field and use, for example the baryon features in the two-point correlation function or power spectrum to constrain dark energy and the expansion history of the universe \cite{McDonald:2006qs}. While we have focused on the power spectrum gotten from 1D measurements of the flux fluctuation, the analysis could be extended to include correlations between different quasar spectra (and therefore different sight-lines). Additional corrections due to correlations between flux and convergence across different lines of sight ($\sim \langle\delta_{f}(\chi\thetaB)\delta_{f}(\chi'\thetaB')\kappa(\chi\thetaB)\rangle$) would arise. However, since lensing is strongest for lenses along the same line of sight as the sources, the terms calculated in this work should be dominant.  One would therefore expect the lensing bias to the correlation function around the baryon scale to remain $0.1-1\%$ \footnote{We have checked that unlike the case of the 3D galaxy correlation function \cite{Hui:2007cu,Hui:2007tm} nothing is remarkably different about the lensing bias to the real space flux correlation function as compared to Fourier space.}. 

\acknowledgements{We are extremely grateful to Lars Hernquist and Volker Springel for providing us with the simulations data used in our analyses.  We thank David Hogg, Guinevere Kauffman, Ian McGreer, Uros Seljak, David Spergel and David Weinberg for discussions. L.H. thanks members of the CCPP (New York), CITA (Toronto) and IAS (Princeton) for their fabulous hospitality. M.L. and L.H. acknowledge support from the DOE under DE-FG02-92ER40699, the
NASA ATP under 09-ATP09-0049, and the Initiatives in Science and Engineering Program at Columbia University. S.M. is supported by the Cyprus State Scholarship Foundation. M.L. is supported as a Friends of the Institute for Advanced Study Member.}

\appendix

\section{Derivation of Magnitude, Number and Observable Cross-correlations}
\label{derive}
\begin{figure}
\begin{center}
$\begin{array}{cc}
\includegraphics[width=0.48\textwidth]{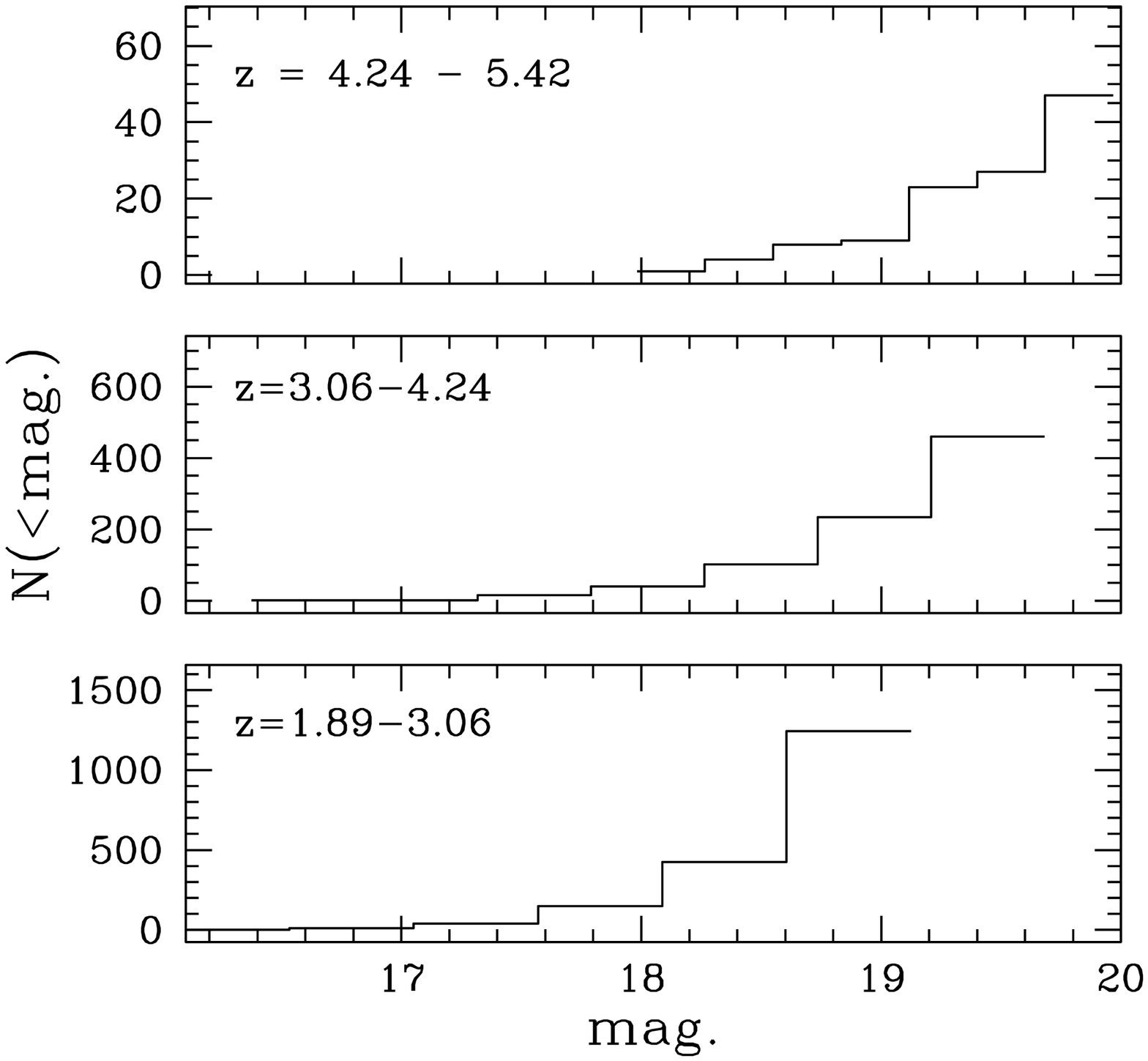}&\includegraphics[width=0.48\textwidth]{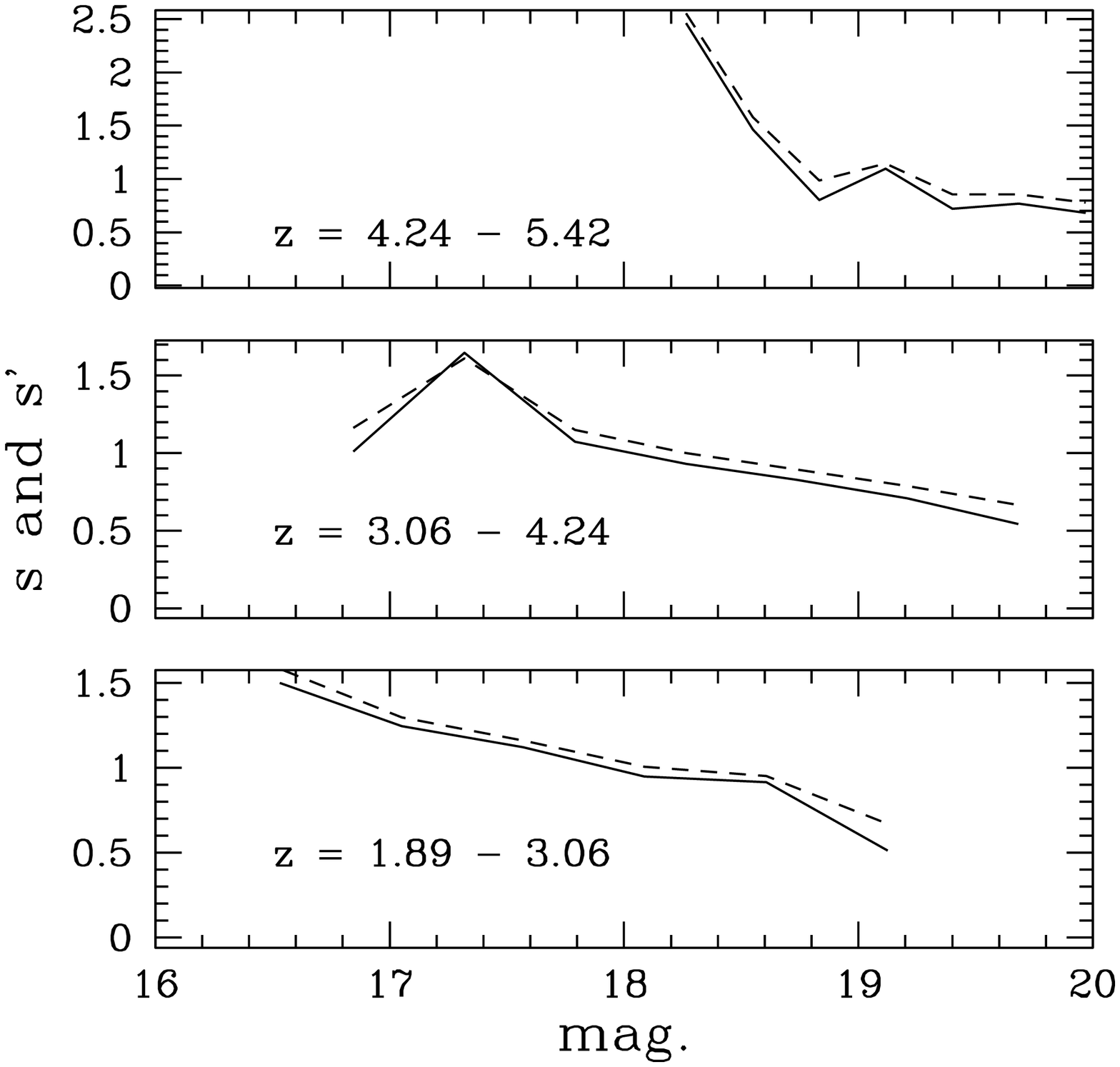}\\
\mbox{(a)}&\mbox{(b)}
\end{array}$
\caption{\label{Fig:slope}(a) A histogram of the number of quasars
as function of $i$-band magnitude measured from the Sloan Digital Sky Survey. 
$(b)$ The slopes $s$ (solid) and $s'$ (dashed) defined in Eqs. (\ref{nlensed}) 
and (\ref{sprime}) as a function of the limiting magnitude
(i.e. assuming a step function $\epsilon$). These measurements ignore
possible incompleteness in the data, and are therefore likely underestimates,
especially at the faint end.}
\end{center}
\end{figure}

In this appendix, we derive the main results of this paper in a unified manner.
They include the lensing induced measurement bias (for both uniform
and flux weighting) and the general magnitude-observable cross-correlation,
described respectively in Eqs. (\ref{nlensed}), 
(\ref{Eq:MagBiasFinal}), (\ref{Eq:MagBiasFluxWeight})
\& (\ref{Eaverage}). Most of these reduce to known results in the literature
for a step-function selection. 

All quantities of interest take the following form:
\begin{eqnarray}
\label{Qdef}
Q = {\sum_I w_I \mathcal{O}_I \over \sum_I u_I}
\end{eqnarray}
where $I$ sums over quasars in one's sample, $w_I$ and $u_I$ are explicit
weights one might apply to them ($w_I$ and $u_I$ might or might not be equal),
and $\mathcal{O}_I$ is a Lyman-alpha forest observable associated with quasar $I$.

It is helpful following \S \ref{LensBiasDeriv} to conceptually pixelize
the survey (with pixel label $i$), and rewrite this as
\begin{eqnarray}
\label{Qdef2}
Q = {\sum_i \int dm \, w(m) \epsilon(m) n_i (m) \mathcal{O}_i
\over \sum_i \int dm \, u(m) \epsilon(m) n_i (m)}
\end{eqnarray}
where $dm \, n_i(m)$ is the number density of quasars at
pixel $i$ with magnitude $m \pm dm/2$, $\epsilon(m)$ is the selection function
(e.g. $\epsilon(m)$ could be $1$ for all quasars brighter than some
limit $m_{\rm lim.}$ and $0$ otherwise), and we assume the weights
$w$ and $u$ are functions of the observed quasar magnitude $m$.

Recall that lensing modifies the observe flux by
$f \rightarrow f + \delta f = f \mu = f (1 + 2\kappa)$,
where $\kappa$ is the (weak) lensing convergence defined previously in 
Eq. (\ref{Eq:KappaDef}). Flux and magnitude are related by
$f = {\,\rm exp\,} [-m {\,\rm ln\,} 10 /2.5]$, and so
$m = m^0 + \delta m = m^0 - {5 \over {\rm ln\,} 10}\kappa$, where
$m^0$ and $m$ are the unlensed and lensed magnitudes respectively
(Eq. [\ref{deltam}]). The observed number density of quasars $n_i (m) dm$ is
related to the pre-lensed number density $n^0_i (m^0) dm^0$ by
\begin{eqnarray}
n_i (m) dm = n^0_i (m^0) dm^0 {1\over 1 + 2\kappa_i}
\end{eqnarray}
where $1/(1 + 2\kappa_i)$ factor accounts for the geometrical increase in area
by magnification (and therefore reduction in number density).
Taylor expanding gives us
\begin{eqnarray}
n_i (m) = n^0_i (m) + \left( {2.5 \over {\rm ln\,} 10} {dn^0_i (m)\over dm} - n^0_i(m)\right)
2 \kappa_i
\end{eqnarray}

The denominator of $Q$ is well approximated by
\begin{eqnarray}
{\rm denominator \,\, of \,\,} Q = \sum_i \int dm \, u(m) \epsilon(m) n^0(m)
\end{eqnarray}
where we have assumed that the survey is large enough that under the summation
over pixels $i$, $n^0_i (m)$ which could fluctuate from pixel to pixel can
be replaced by its average $n^0 (m)$ (i.e. the true pre-lensed luminosity function),
and $\sum_i \kappa_i \sim 0$.
The summation $\sum_i$ then simply gives the number of pixels in the survey. 
This kind of approximation is equivalent to ignoring corrections of the
integral constraint type \cite{Hui:1998ix}.

The numerator of $Q$ can be computed by rewriting $\mathcal{O}_i = \mathcal{O}_{\rm true}
+ \delta \mathcal{O}_i$, where $\mathcal{O}_{\rm true}$ denotes the ensemble 
average $\mathcal{O}_{\rm true} = \langle \mathcal{O}_i\rangle$ (i.e. an average over
ensemble of realizations of the universe). 
Assuming there is neither correlation between $\mathcal{O}_i$ and the pre-lensed number
density $n^0_i$, nor correlation between $\kappa_i$ and $n^0_i$, we find
\begin{eqnarray}
\langle {\rm numerator \,\, of \,\,}Q \rangle = \sum_i \int dm \, w(m) \epsilon(m) n^0(m) \mathcal{O}_{\rm true} + \sum_i \int dm \, w(m) \epsilon(m) \left( {2.5 \over {\rm ln\,} 10} {dn^0 (m)\over dm} - n^0(m)\right) 2 \langle \kappa \delta\mathcal{O} \rangle
\end{eqnarray}
where we have dropped the $i$ label from $\langle \kappa_i \delta\mathcal{O}_i \rangle$ since this is simply a cross-correlation at zero-lag. Just as for the denominator,
the summation $\sum_i$ reduces simply to the total number of pixels.

In summary, we find
\begin{eqnarray}
\label{fundamental}
\langle Q \rangle = \mathcal{O}_{\rm true} 
{\int dm \, w(m) \epsilon(m) n^0(m) \over \int dm \, u(m) \epsilon(m) n^0(m)}
+ 2\langle \kappa \delta\mathcal{O} \rangle
{\int dm \, w(m) \epsilon(m) \left( {2.5 \over {\rm ln\,} 10} {dn^0 (m)\over dm} - n^0(m)\right) \over \int dm \, u(m) \epsilon(m) n^0(m)}
\end{eqnarray}
This is a fundamental result from which everything follows.
Let's first apply it to the case with $w(m) = u(m) = 1$ i.e. $Q$ is what we
have been calling $\mathcal{O}_{\rm obs}$, with equal weights applied to
all quasars {\it within} one's sample. 
Eq. (\ref{fundamental}) tells us
\begin{eqnarray}
\label{AP1}
\langle \mathcal{O}_{\rm obs} \rangle = \mathcal{O}_{\rm true} + (5s-2) \langle \kappa \delta\mathcal{O}
\rangle \quad , \quad s = {1\over {\rm ln\,} 10} {\int dm \, \epsilon(m) (dn^0/dm) \over
\int dm \, \epsilon(m) n^0(m)} \, ,
\end{eqnarray}
consistent with Eq. (\ref{Eq:MagBiasFinal}) and the definition of $s$ in
Eq. (\ref{nlensed}). Note that $\langle \kappa \delta\mathcal{O} \rangle
= \langle \kappa \mathcal{O} \rangle$ since $\langle \kappa \rangle = 0$. 
If the selection function $\epsilon(m)$ were a
step-function, $s$ reduces to the more familiar slope of the quasar number count
upon integration by parts (see \S \ref{Intro}). 

Let's next try $w(m) = u(m) = {\rm exp\,} [-m {\,\rm ln\,}10/2.5]$ in Eq. (\ref{fundamental}),
i.e. $Q$ is equivalent to $\mathcal{O}_{\rm obs}$, with a flux weighting.
It is simple to see that 
\begin{eqnarray}
\label{AP2}
\langle \mathcal{O}_{\rm obs} \rangle
= \mathcal{O}_{\rm true} + (5s'-2) \langle \kappa \delta\mathcal{O} \rangle
\quad , \quad s' = {1\over {\rm ln\,} 10} {\int dm \, \epsilon(m) (dn^0 /dm)
10^{-m/2.5}
\over \int dm \, \epsilon(m) n^0(m) 10^{-m/2.5}}
\end{eqnarray}
which is consistent with Eqs. (\ref{Eq:MagBiasFluxWeight}) \& (\ref{sprime}).

Lastly, let's use $w(m) = m - \bar m$ and $u(m) = 1$ in 
Eq. (\ref{fundamental}), with $\bar m = \int dm \, m \, \epsilon(m) 
n^0 (m) / \int dm\, \epsilon(m) n^0 (m)$. Then, $Q$ corresponds to
the cross-correlation estimator $\mathcal{E}$ of Eq. (\ref{EstimatorE}),
keeping in mind that $\bar m$ (which is the average sample magnitude defined by
the {\it true} unlensed luminosity function) should be well approximated
by $\sum_I m_I / \sum_I$ for a sufficiently large survey.
We obtain:
\begin{eqnarray}
\label{AP3}
\langle \mathcal{E} \rangle = 5 \tilde s
\langle \kappa \delta\mathcal{O} \rangle \quad , \quad
\tilde s \equiv {1\over {\rm ln\,} 10} 
{\int dm \, \epsilon(m) (dn^0 /dm) (m - \bar m) 
\over \int dm \, \epsilon(m) n^0(m)}
\end{eqnarray}
which is consistent with Eqs. (\ref{Eaverage}) \& (\ref{stilde}).
Note how the `$-2$' that is present in both
Eqs. (\ref{AP1}) \& (\ref{AP2}) is absent in (\ref{AP3}).
In the first two cases, this `$-2$' originates from
the `$- n^0(m)$' in the second term of Eq. (\ref{fundamental}).
This term yields zero in Eq. (\ref{AP3}) by definition
of $w(m) = m - \bar m$. 

The different symbols $s$, $s'$ and $\tilde s$ can be seen as different (normalized) moments of $\epsilon(m)dn^0/dm$. In Fig. \ref{Fig:slope} we show the cumulated number counts and rough estimates of $s$ and $s'$ from SDSS data release $6$. It is worth emphasizing that incompleteness is not taken into account
in deducing these estimates, and the true values of $s$ and $s'$ could well be higher, especially at the faint end. In other words, if incompleteness is
present, it should be properly taken into account through the efficiency $\epsilon(m)$. Blindly measuring $s$ by taking derivative of the {\it observed} (i.e. incomplete) cumulated number count would result in an underestimate.

\section{Determining the Low-$k$ Bispectrum}
\label{Extrap}
\begin{figure}
\begin{center}
\includegraphics[width=0.45\textwidth]{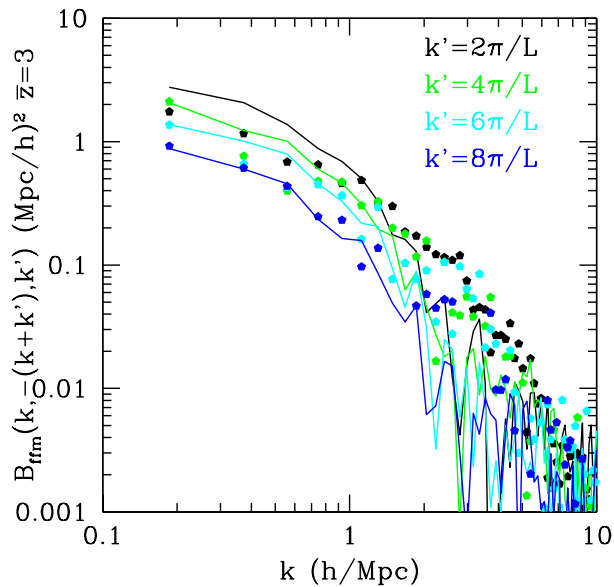}
\caption{\label{Fig:Hierarchical}The flux-flux-mass bispectrum (points) compared with the hierarchical model given in Eq. (\protect\ref{Eq:Hierarch}) (solid lines) at redshift $3$. The simulations box-size $L=33.75\MpcOh$.}
\end{center}
\end{figure}

In this Appendix we discuss how we determine the flux-flux-mass bispectrum at low-$k_{||}$. The issue is that the simulations have a finite box size and we need to extrapolate $B_{ffm}(k_{||},-(k_{||}+k_{||}'),k_{||}')$ from $k_{||}'\sim 2\pi/L$ to $k_{||}'\sim 0$, where $L$ is the size of the box. At $z=3$ we have simulations in a larger box and we find that doubling the box size (from $L=50 \MpcOh$ to $L=100 \MpcOh$) increases $B_{ffm}(k_{||},-k_{||},0)$ by a factor of about four. To determine $B_{ffm}$ for very low $k$ we first use a hierarchical model for $B_{ffm}$, 
\be
\label{Eq:Hierarch}
B_{ffm}(k_{||},-(k_{||}+k_{||}'),k_{||}')=Q_{ffm}\left(P_{fm}(k_{||})P_{fm}(k_{||}')+P_{fm}(k_{||})P_{fm}(k_{||}+k_{||}')+P_{fm}(k_{||}')P_{fm}(k_{||}+k_{||}')\right)\, .
\ee
This hierarchical model with $Q_{ffm}$ independent of $k_{||}$ but varying with $z$ appears to be a pretty good fit to the simulations-measured bispectrum for $k_{||},k_{||}'\lsim 1\hOMpc$ with better agreement $z=3$ and $z=4$ than $z=2$. At $k_{||} \gsim 1\hOMpc$ the hierarchical bispectrum drops off much more rapidly than the true bispectrum. Comparison of the hierarchical and true bispectra is shown in Fig. \ref{Fig:Hierarchical}. Given the agreement between the hierarchical and the measured bispectra (on large scales), we now assume Eq. (\ref{Eq:Hierarch}) works at $B_{ffm}(k_{||},-k_{||},0)$ and embark on the somewhat easier task of determining $P_{fm}$ at very low $k_{||}$. 

The low $k_{||}$ value of $P_{fm}$ is determined using both the simulations and analytics. Precisely, we set $P_{fm}(k=0)=P^{sims}_{fm}(k_{min})/P^{calc.}_{bm}(k_{min})\times P^{calc.}_{bm}(k=0)$. The calculated 1-D baryon-mass power spectrum at $k_{||}=0$ is given by
\be
P_{bm}^{calc.}=\int \frac{d^2{\bf k}_\perp}{(2\pi)^2}P_{3D}(k_{\perp})e^{-k_\perp^2/k_s^2}
\ee
where $P_{3D}(k)$ is the 3-D mass power spectrum, $k_s=\sqrt{\frac{10}{3}}k_J$ with $k_J=\sqrt{4\pi G\rho}/(c_s(1+z))$ is the Jeans length for a system with sounds speed $c_s$ \cite{Gnedin:1997td} and proper mass density $\rho$. Assuming the intergalactic medium can be treated as a monatomic ideal gas $c_s=\sqrt{5T/(3m)}$ where $m$ is the mean particle mass. For a fully ionized gas composed of $75\%$ Hydrogen and $25\%$ Helium by mass $m=0.59m_p$  where $m_p$ is the proton mass.  We assume $T=20,000 K$ is the gas temperature constant over the redshift range we are interested in \cite{Zaldarriaga:2000mz}. 

The reader might wonder whether analytic estimates of the flux-mass power spectrum and flux-flux-mass bispectrum can be used to make accurate predictions without simulations. A naive guess,  $\delta_f=b\,\delta_b(k)$ works fairly well for $P_{fm}$ but overestimates the magnitude of $B_{ffm}$ by a factor of a few. The dominant reason for this discrepancy appears to be the difference between $Q_{ffm}$ and $Q_{mmm}$, where $Q_{mmm}$ is the equivalent parameter for the hierarchical mass-mass-mass bispectrum.  Hyper-extended perturbation theory \cite{Scoccimarro:1998uz} suggests that on the scales we are interested in, the hierarchical ansatz should fit the mass bispectrum $B_{mmm}$ with a factor $Q_{mmm}\sim 3$ (and only weakly dependent on redshift). Indeed our simulations give a similar value for $Q_{mmm}$, and the hierarchical ansatz fits $B_{mmm}$ rather well. However, for the flux-flux-mass bispectrum we find that $Q_{ffm} \lsim 1 $ and varies more strongly with redshift.  So, because of this difference the assumption $B_{ffm}(k,k',k'')=b^2B_{mmm}$ fails even on large-scales where $P_{ff}\sim b^2P_{mm}$. This is why we have to rely on simulations to obtain an estimate of the flux-flux-mass bispectrum, but making suitable corrections to get
at the $(k_\parallel, -k_\parallel, 0)$ limit. It is also worth emphasizing we have estimated the measurement bias
associated with the flux power spectrum two different ways, one using
the flux-flux-mass bispectrum and the other doesn't, and they agree
reasonably well (Fig. \ref{Fig:PffBispec}).

\section{Estimator and Error for the Magnitude-Flux Correlation Amplitude}
\label{estimator}

Recall that the expected magnitude-flux correlation
from Eq. (\ref{magfluxexpect}) takes the following form
\begin{eqnarray}
\langle \mathcal{E}_{\delta m \, \delta_f} (\chi) \rangle
= g(\chi) A
\end{eqnarray}
where $g(\chi)$ is a function of $\chi$ (distance to absorption), and
$A$ represents the amplitude of this correlation, which one can equate
with $P_{fm} (k_\parallel = 0)$ if one wishes (the precise choice
will have no bearing on the final signal-to-noise of interest).
We have suppressed the
dependence on $\chi_Q$ (distance to quasar).

Our goal is to come up with an estimator for this amplitude $A$
by combining the magnitude-flux correlations at different scales ($\chi$'s),
and show that its error bar is given by Eq. (\ref{magfluxS2N}).
The estimator takes the form:
\begin{eqnarray}
\mathcal{E}_A = \sum_{I, \chi} w^I(\chi) \delta m^I \delta_f^I (\chi)
\end{eqnarray}
where the $I$ index labels the quasar, and the sum over $\chi$ ranges
over the (binned) scales of Lyman-alpha forest observations.
Here, $\delta m^I = m^I - \bar m$, with $\bar m$ being the mean
sample magnitude (see discussion around Eq. [\ref{AP3}]), and
$w^I(\chi)$ represents a weighting of the data over quasars and absorption
locations which remains to be specified. To satisfy
$\langle \mathcal{E}_A \rangle = A$, we would want $w^I(\chi)$ to satisfy
\begin{eqnarray}
\label{wIchinorm}
\sum_{I,\chi} w^I(\chi) g(\chi) = 1
\end{eqnarray}

The variance of this estimator is given by
\begin{eqnarray}
\langle \delta \mathcal{E}_A^2 \rangle
= \sum_{I, J, \chi, \chi'} w^I(\chi) w^J(\chi')
\left[
\langle \delta m^I \delta^I_f(\chi) \delta m^J \delta^J_f (\chi')
\rangle - \langle \delta m^I \delta^I_f(\chi) \rangle
\langle \delta m^J \delta^J_f (\chi') \rangle \right]
\end{eqnarray}
Assuming the terms in $\,\left[\,\,\right]\,$ are dominated by
$\delta_{IJ} \langle \delta m^2 \rangle 
\langle \delta_f (\chi) \delta_f (\chi') \rangle$
(i.e. correlations between different quasars are weak, and that
cross correlations between magnitude and $\delta_f$ are also weak
compared with auto correlations), we find
\begin{eqnarray}
\label{dEA2}
\langle \delta \mathcal{E}_A^2 \rangle
\sim 
\langle \delta m^2 \rangle \sum_{I, \chi, \chi'} w^I(\chi)
\xi_{ff} (\chi, \chi') w^I(\chi')
\end{eqnarray}
where we have used $\xi_{ff} (\chi, \chi')$ to represent
$\langle \delta_f (\chi) \delta_f (\chi') \rangle$ -- it should be kept
in mind that this should include contributions from both the intrinsic
forest fluctuations
and shot-noise. The shot-noise contribution makes $\xi_{ff}$ strictly
speaking a function of the quasar index $I$, but we will ignore
it for our purpose of a crude $S/N$ estimate.

Standard minimization technique applied to Eq. (\ref{dEA2}) subject
to the constraint Eq. (\ref{wIchinorm}) gives us the minimum variance weighting:
\begin{eqnarray}
w^I(\chi) = \left[ \sum_{\chi'} \xi_{ff}^{-1} (\chi, \chi') g(\chi') \right] /
\left[ \sum_{J, \chi', \chi''} g(\chi') \xi_{ff}^{-1} (\chi', \chi'') g(\chi'')
\right]
\end{eqnarray}
where $\xi_{ff}^{-1}$ is defined to be the matrix inverse of
$\xi_{ff}$ i.e. $\sum_{\chi'}
\xi_{ff}^{-1}(\chi, \chi') \xi_{ff}(\chi', \chi'')
= \delta_{\chi, \chi''}$.
With this optimal weighting, the variance is given by
\begin{eqnarray}
\langle \delta \mathcal{E}_A^2 \rangle
= {\langle \delta m^2 \rangle \over N_{\rm QSO}}
\left[ \sum_{\chi', \chi''} g(\chi') \xi_{ff}^{-1} (\chi', \chi'')
g(\chi'') \right]^{-1}
\end{eqnarray}
where $N_{\rm QSO}$ is the number of quasars.

The signal-to-noise of interest is
\begin{eqnarray}
{S \over N} = 
{A \over \sqrt{\delta \mathcal{E}_A^2 \rangle}}
\end{eqnarray}
which after some Fourier manipulations is equivalent to 
Eq. (\ref{magfluxS2N}).

Let us close by noting that in the same spirit and notation, the
magnitude-flux correlation (at a given $\chi$) has a variance
of 
\begin{eqnarray}
{\langle \delta m^2 \rangle \over N_{\rm QSO}}
\xi_{ff} (\chi, \chi)
\end{eqnarray}

\bibliography{LyAlpha}

\end{document}